\documentclass[sigconf,natbib=true]{acmart}

\usepackage{amsmath,amsfonts}
\usepackage{adjustbox}
\usepackage{algorithm}
\usepackage{algcompatible}
\usepackage{algpseudocode}
\usepackage{bm}
\usepackage{graphicx}
\usepackage{textcomp}
\usepackage{xcolor}
\usepackage{tabularx}
\usepackage{multirow}
\usepackage{makecell}
\usepackage{diagbox}

\AtBeginDocument{
  \providecommand\BibTeX{{
    \normalfont B\kern-0.5em{\scshape i\kern-0.25em b}\kern-0.8em\TeX}}}

\setcopyright{acmcopyright}
\copyrightyear{2022}
\acmYear{2022}

\author{Yoonhyuk Choi}
\orcid{0000-0003-4359-5596}
\affiliation{
  \institution{
  Seoul National University}
  \city{Seoul}
  \state{Republic of Korea}
}
\email{younhyuk95 @ snu.ac.kr}

\author{Jiho Choi}
\orcid{0000-0002-7140-7962}
\affiliation{
  \institution{
  Seoul National University}
  \city{Seoul}
  \state{Republic of Korea}
}
\email{jihochoi @ snu.ac.kr}

\author{Taewook Ko}
\orcid{0000-0001-7248-4751}
\affiliation{
  \institution{
  Seoul National University}
  \city{Seoul}
  \state{Republic of Korea}
}
\email{taewook.ko @ snu.ac.kr}

\author{Hyungho Byun}
\orcid{0000-0003-1908-637X}
\affiliation{
  \institution{
  Seoul National University}
  \city{Seoul}
  \state{Republic of Korea}
}
\email{notorioush2 @ snu.ac.kr}

\author{Chong-Kwon Kim}
\orcid{0000-0002-9492-6546}
\affiliation{
  \institution{Korea Institute of Energy Technology}
  \city{Naju}
  \state{Republic of Korea}
}
\email{ckim @ kentech.ac.kr}

\renewcommand{\shortauthors}{Yoonhyuk Choi et al.}

\acmConference[CIKM '22] {Proceedings of the 31st ACM International Conference on Information and Knowledge Management}{October 17--21, 2022}{Atlanta, GA, USA.}
\acmBooktitle{Proceedings of the 31st ACM International Conference on Information and Knowledge Management (CIKM '22), October 17--21, 2022, Atlanta, GA, USA}
\acmPrice{15.00}
\acmISBN{978-1-4503-9236-5/22/10}
\acmDOI{10.1145/3511808.3557434}

\begin{document}

\title{Review-Based Domain Disentanglement without Duplicate Users or Contexts for Cross-Domain Recommendation}

\begin{abstract} 
A cross-domain recommendation has shown promising results in solving data-sparsity and cold-start problems. Despite such progress, existing methods focus on domain-shareable information (overlapped users or same contexts) for a knowledge transfer, and they fail to generalize well without such requirements. To deal with these problems, we suggest utilizing review texts that are general to most e-commerce systems. Our model (named \textit{SER}) uses three text analysis modules, guided by a single domain discriminator for disentangled representation learning. Here, we suggest a novel optimization strategy that can enhance the quality of domain disentanglement, and also debilitates detrimental information of a source domain. Also, we extend the encoding network from a single to multiple domains, which has proven to be powerful for review-based recommender systems.
Extensive experiments and ablation studies demonstrate that our method is efficient, robust, and scalable compared to the state-of-the-art single and cross-domain recommendation methods.
\end{abstract}



\maketitle

\section{Introduction}
With the rapid growth of e-commerce, recommender systems have become an obligatory tool for interconnecting customers with relevant items. Early schemes suffer from cold-start problems caused by data insufficiency. To tackle this problem, auxiliary information of data such as social relations \cite{fu2021dual}, the trustworthiness of reviewers \cite{banerjee2017whose}, item images \cite{yu2018aesthetic} are exploited. Especially, textual or linguistic information such as reviews are commonly available, and many text-aided recommendation algorithms have been introduced \cite{zheng2017joint,chen2018neural,chen2019co,dong2020asymmetrical,wang2021leveraging,shi2021wg4rec,xiong2021counterfactual,shuai2022review}.
For example, some studies infer the preferences of users by applying natural language processing (NLP) techniques such as topic modeling \cite{mcauley2013hidden,bao2014topicmf}. Recently, some apply convolutional neural network (CNN) on textual reviews \cite{zheng2017joint}, others \cite{chen2018neural,dong2020asymmetrical} further utilize attention mechanism to deal with important reviews. Many text analysis modules are considerable, but here, we employ a simple text analysis module \cite{zheng2017joint} to focus on a cross-domain recommendation scenario.

Along with the text-based recommender systems, numerous cross-domain recommendation \cite{hu2018conet,yuan2019darec,fu2019deeply} and transfer learning \cite{yuan2020parameter,yang2021autoft,hande2021domain} approaches have been introduced. They leverage the information learned from source domains to improve the quality of recommendation in a target domain. Some of them suggest a fine-tuning network, but they may suffer from catastrophic forgetting \cite{chen2019catastrophic}. Context mappings \cite{yuan2019darec,li2020ddtcdr,liu2021collaborative,guo2021gcn} are another branch of transfer learning, which map shareable information from a source to the target domain. However, these approaches require the same contexts or overlapped users, which may restrict their applicability severely \cite{kang2019semi}. In a real-world dataset, this kind of information might be scarce, and thus, their contributions might be insignificant, which still requires further investigation. 

To solve such constraints, we focus on another branch of methods that are independent of specific users or contexts \cite{cai2019learning,zhao2020catn,krishnan2020transfer}. However, it is still challenging to achieve knowledge transfer between heterogeneous domains (e.g., totally different users or items), while debilitating noises.
For one solution, domain adaptation (DA) \cite{ben2010theory,ganin2016domain,li2020maximum,li2021learning} minimizes source risk as well as H-divergence, capturing domain-indiscriminative information. Nonetheless, these methods are highly dependent on the consistency (maximum mean discrepancy) between domains. Though disentangled representation learning algorithms \cite{gretton2005measuring,bousmalis2016domain,li2019learning,peng2019domain} concurrently extract domain-specific, and domain-invariant knowledge to distill pertinent knowledge from multiple seemingly counterproductive domains, they also suffer from measuring theoretical bound and tractability of a log-partition function \cite{gabrie2019entropy,mcallester2020formal}.

In this paper, we suggest a novel method of disentangled representation learning that utilizes textual reviews common to most e-commerce systems. Here, we focus on debilitating noises of another domain and claim that this can be achieved by reinforcing the discrimination power of a domain discriminator. Specifically,
using the connections between mutual information (MI) estimation and DA \cite{kang2019contrastive,saito2019semi,park2020joint,zhao2021domain}, we integrate the role of the MI estimator with a single domain discriminator. Thus, the domain-shareable information is further utilized as an input for the domain discriminator, which can reinforce the discrimination power. In terms of adversarial training, a well-trained discriminator guides three types of FEs to secure robustness and improve the recommendation quality for both domains.
We perform extensive experiments on the publicly available dataset to compare our model under single and cross-domain recommendation scenarios. The quantitative and qualitative analysis demonstrates the superiority of our suggestions. The contributions can be summarized as follows.
\begin{itemize}
\item We propose $SER$ that adaptively disentangles features depending on the characteristics of the source and target domains, where the novel optimization strategy leads to a promising solution for a recommendation. Consequently, our approach is also applicable to heterogeneous scenarios, in case the source and target domain have less similar characteristics (e.g., non-overlapping users or contexts).
\item Our approach is comprehensive that is closely integrated with text-based feature extraction. Unlike previous cross-domain recommendation schemes that require duplicate entities from heterogeneous domains, we focus on retrieving review information that is common for most domains. 
\item We perform extensive experiments to answer the important research questions described in Section \ref{experiments}. The results indicate the superiority and robustness of our proposed method.
\end{itemize}

\section{Related Work}
In this section, we describe an overview of existing recommender systems; text-aided single-domain recommendations, extending to a cross-domain scenario, and applying domain adaptation to achieve disentangled representation learning, which can enhance the knowledge transfer between domains.

\subsection{Text-aided Recommender System}
Textual information is the most popular side information and many text-based methods \cite{zheng2017joint,chen2019dynamic,chen2019co,dong2020asymmetrical} have been proposed recently. Previous techniques simply integrate DNN-based feature extraction with MF for a rating prediction, while \cite{zheng2017joint} utilizes two parallel CNNs, and \cite{li2017neural} adopts Gated Recurrent Units for review analysis. Attention mechanisms are widely used also to pinpoint useful words and reviews \cite{seo2017interpretable,chen2018neural,tay2018multi,dong2020asymmetrical}. 
Even though prior works show the usefulness of textual information, the limitation of review information due to the limited size of training data, and the irrelevance of reviews toward target items have been raised also \cite{sachdeva2020useful,zeng2021zero}. To tackle this problem, we now introduce recent algorithms for cross-domain recommender systems which adopt supplementary domains.

\subsection{Cross-Domain Recommendation (CDR)}
CDR utilizes information from source domains to alleviate the cold-start problem in the target domain.
Especially, most of them assume embedding-based or rating pattern-based transfer scenario \cite{zhu2021cross}.
In detail, early studies \cite{elkahky2015multi,man2017cross} adopt a feature mapping technique that requires overlapped users. For example, RC-DFM \cite{fu2019deeply} applies Stacked Denoising Autoencoder (SDAE) to each domain, where the learned knowledge of the same set of users is transferred from a source to the target domain. More recently, to overcome the requirements of overlapped users, there was an attempt \cite{wang2018cross,zhao2020catn} that employs similar users for feature mapping. However, they implicate the limitation of debilitating noises from a source domain. In this paper, we adopt review texts that are common to most e-commerce systems without requiring overlapping users or contexts. Further, we suggest a novel disentangled representation learning for a knowledge transfer between domains which can reduce noises from a source domain. We now introduce some traditional
methods for disentangled representation learning below.

\subsection{Disentangled Representation Learning}
Recently, many efforts have been dedicated to capturing domain-shareable information. Especially, with the powerful mechanism of adversarial training, DA has been adopted for various fields; question answering tasks \cite{ramakrishnan2018overcoming}, and recommendation scenarios \cite{yuan2019darec,bonab2021cross}. Some adopt textual reviews under no user or context overlap for a cross-domain recommendation \cite{yu2020semi}. However, these approaches only focus on domain-shareable knowledge, while ignoring domain-specific ones. MMT \cite{krishnan2020transfer} further captures domain-specific knowledge, but the disentanglement between domain-common features is not applied. Though DADA \cite{peng2019domain} introduced domain-agnostic learning, the domain discriminator is only utilized for the extraction of a domain-common feature. Since the domain-specific knowledge is solely guided using MI minimization \cite{belghazi2018mutual}, they have shown to implicate some defects \cite{mcallester2020formal}. In this paper, we clear the aforementioned limitations, adopting a framework for the simultaneous extraction of domain-specific and domain-common knowledge by integrating MI estimator with domain discriminator. Consequently, our model enhances the quality of disentangled representation learning without requiring overlapping users or contexts for a recommendation.

\begin{figure*}[ht]
  \includegraphics[width=\textwidth]{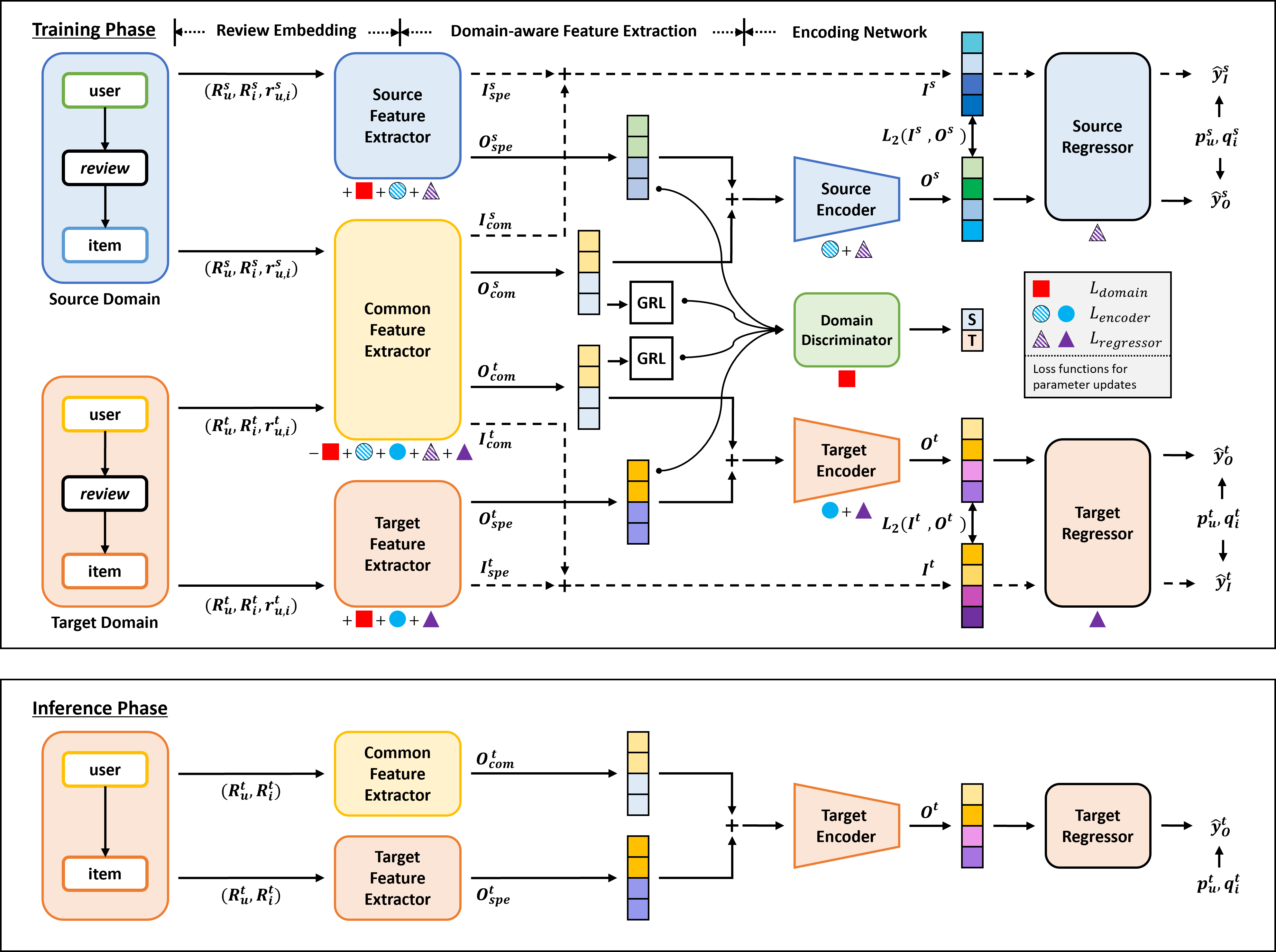}
  \caption{The overall framework of SER for a cross-domain recommendation scenario}
  \label{model}
\end{figure*}

\begin{table}[htbp]\caption{Notations}
\vspace{-1.5em}
\begin{center}
     \begin{tabular}{@{}l|l@{}}
\toprule
\multicolumn{1}{c|}{\textbf{Symbol}} & \textbf{Explanation} \\
\toprule
\multicolumn{1}{c|}{$D^s,D^t$} & Set of source and target domain dataset\\
\multicolumn{1}{c|}{$u,i,y_{u,i},r_{u,i}$} & Set of user, item, rating, and individual review\\
\multicolumn{1}{c|}{$R_u,R_i$} & Set of aggregated reviews for user and item \\
\multicolumn{1}{c|}{$p_u, q_i$} & Rating embedding of user and item  \\
\multicolumn{1}{c|}{$O$} & Extracted feature of $R_u,R_i$ \\
\multicolumn{1}{c|}{$I$} & Extracted feature of $r_{u,i}$ \\
\multicolumn{1}{c|}{$\widehat{d}$} & Predicted domain label of $O$ \\
\multicolumn{1}{c|}{$\widehat{y}_E,\widehat{y}_I$} & Predicted label of $E$ and $I$ \\
\multicolumn{1}{c|}{MI} & Mutual information \\
\multicolumn{1}{c|}{$N_s,N_t$} & Mini-batch training samples of $s$ and $t$ \\
\multicolumn{1}{c|}{$\phi$} & Word embedding function \\
\multicolumn{1}{c|}{$\oplus$} & Concatenation operator \\
\multicolumn{1}{c|}{$\mathcal{L}$} & Loss function \\
\bottomrule
    \end{tabular}
\end{center}
    \label{notations}
\end{table}

\section{Problem Formulation}
We formulate the cross-domain recommendation task as follows. Assume two datasets, $D^s$ and $D^t$, be the data of the source and target domains. Each of them consists of tuples, $(u, i, y_{u, i}, r_{u,i})$ which represents an individual review $r_{u,i}$ provided by a user $u$ for an item $i$ with rating $y_{u,i}$. 
Our goal is to improve the recommendation quality in a target domain with the aid of a source domain. To achieve this, we suggest a domain-aware knowledge transfer module with encoding networks. The overall notations can be seen in Table \ref{notations}.

\section{Our Approach}

Figure \ref{model} describes our model with the following key components: 

\begin{itemize}
    \item \textbf{Review embedding layer} 
    Vectorize the aggregated and individual review using pre-trained word embedding function, which can be word2vec\footnote{https://code.google.com/archive/p/word2vec} \cite{mikolov2013distributed} or GloVe\footnote{https://nlp.stanford.edu/projects/glove} \cite{pennington2014glove}.
    \item \textbf{Domain-aware feature extraction} Integrated with domain discriminator, three feature extractors (FEs) capture domain-aware knowledge from vectorized aggregated reviews. Especially, the common FE acts as a knowledge transfer module and utilizes both source and target domain data.
    \item \textbf{Encoding network} The encoding networks generate a single vector from vectorized aggregated reviews by aligning them with the latent of an individual review. 
\end{itemize}
We now illustrate the details of each module precisely.

\subsection{\textbf{Review Embedding Layer}}
Given the input data (\textit{u, i, $y_{u,i}$, $r_{u,i}$}), we first ensemble the reviews of user and item. For example, given user $u$, all reviews written by her are treated as a single document $R_u$. The item reviews $R_i$ can be retrieved similarly. However, for each training sample, we exclude an individual review $r_{u,i}$ that $u$ has written after purchasing $i$ since it cannot be used during the inference phase. Here, we do not consider a sequence (or time) of their purchasing histories.

We utilize first $n$ words in $R$ and apply pre-trained word embedding function \cite{mikolov2013distributed,pennington2014glove} for vectorization. The words are mapped to $c$-dimensional vectors, and column-wisely concatenated to form document embedding $V=\phi(w_1) \oplus \phi(w_2) \oplus ...  \oplus \phi(w_n)$, where $\phi$ and $\oplus$ is a embedding and concatenation operation. 

\subsection{\textbf{Domain-aware Feature Extraction}}
As described in Figure \ref{model}, we utilize three types of feature extractors (FEs) to separate the domain-specific and domain-common knowledge. In this paper, each FE adopts the widely used text analysis method \cite{zheng2017joint}. More specifically, using the document embedding $V$, three convolutional $FEs$ produce outputs, followed by a non-linear activation function (e.g., ReLU) with a row-wise max-pooling layer. 
Finally, by concatenating the outputs of stacked convolution layers, we can derive a unified embedding $O$ from user and item reviews (please refer to \cite{zheng2017joint} for text convolutions).
Similarly, we can retrieve the embedding of an individual review $I$ using three FEs.

To summarize, the domain-specific FEs (source and target) generate $O^s_{spe},I^s_{spe}$ and $O^t_{spe},I^t_{spe}$ for each domain. Also, the domain-common FE extracts $O^s_{com},I^s_{com}$, and $O^t_{com},I^t_{com}$ using inputs from source and target domain, respectively. Here, the superscript $s,t$ acts as an identifier of two domains.
Though we assume three FEs, they may contain similar information without any constraints. 
To achieve a better disentanglement, we suggest an optimization strategy from two perspectives: (1) for the extraction of domain-common knowledge, we employ domain adaptation to serve common FE as an intermediate agent for knowledge transfer. (2) for the disentanglement of domain-specific features, we suggest a novel constraint for adversarial training.

\subsubsection{\textbf{Extracting domain-common features}}
For a knowledge transfer between domains, we focus on domain adaptation (DA) \cite{ganin2016domain} that has proven to be effective in case a source domain has a richer label than the target \cite{ben2010theory}.
DA employs domain discriminator with Gradient Reversal Layer (GRL) to reduce the cross-domain discrepancy. The discriminator gives additional penalties (or loss) to common FE for capturing domain-discriminative information, which has proven to be effective for debilitating noises. Here, we adopt two layers of a fully-connected neural network as domain discriminator $F_{disc}$, which takes the output of common FE as below:

\begin{equation} 
\label{domain_label}
\widehat{d}_{com}^s=F_{disc}(O_{com}^s),\, \,\widehat{d}_{com}^t=F_{disc}(O_{com}^t).
\end{equation}
Here, $\widehat{d}_{com}$ stands for predicted domain probability of common feature $O_{com}$, where the error can be calculated through binary cross-entropy loss between true label $d_{com}$ as below:
\begin{equation}
\begin{gathered}
\mathcal{L}_{com}^s= -{1 \over N_s}\sum_{s=1}^{N_s} log(1-\widehat{d}^{s}_{com}),\,\,\mathcal{L}_{com}^t = -{1 \over N_t}\sum_{t=1}^{N_t} log(\widehat{d}^{t}_{com}).
\end{gathered}
\label{d_inv_loss}
\end{equation}
The true label $d_{com}$ is a binary value, \{0, 1\} for source and target domain. Here, $N$ is the mini-batch training samples from two domains. GRL is positioned between common FE and domain discriminator (please refer to Figure \ref{model}), multiplies a negative constant during back-propagation. Consequently, the common FE is reinforced to capture domain-indiscriminative knowledge to fool discriminator.
Nonetheless, DA itself implicates some limitations, which can be sensitive to domain divergence, and prohibitive applicability \cite{chen2019transferability}.  

\begin{figure}
  \centering
  \vspace{-2mm}
  \includegraphics[width=0.47\textwidth]{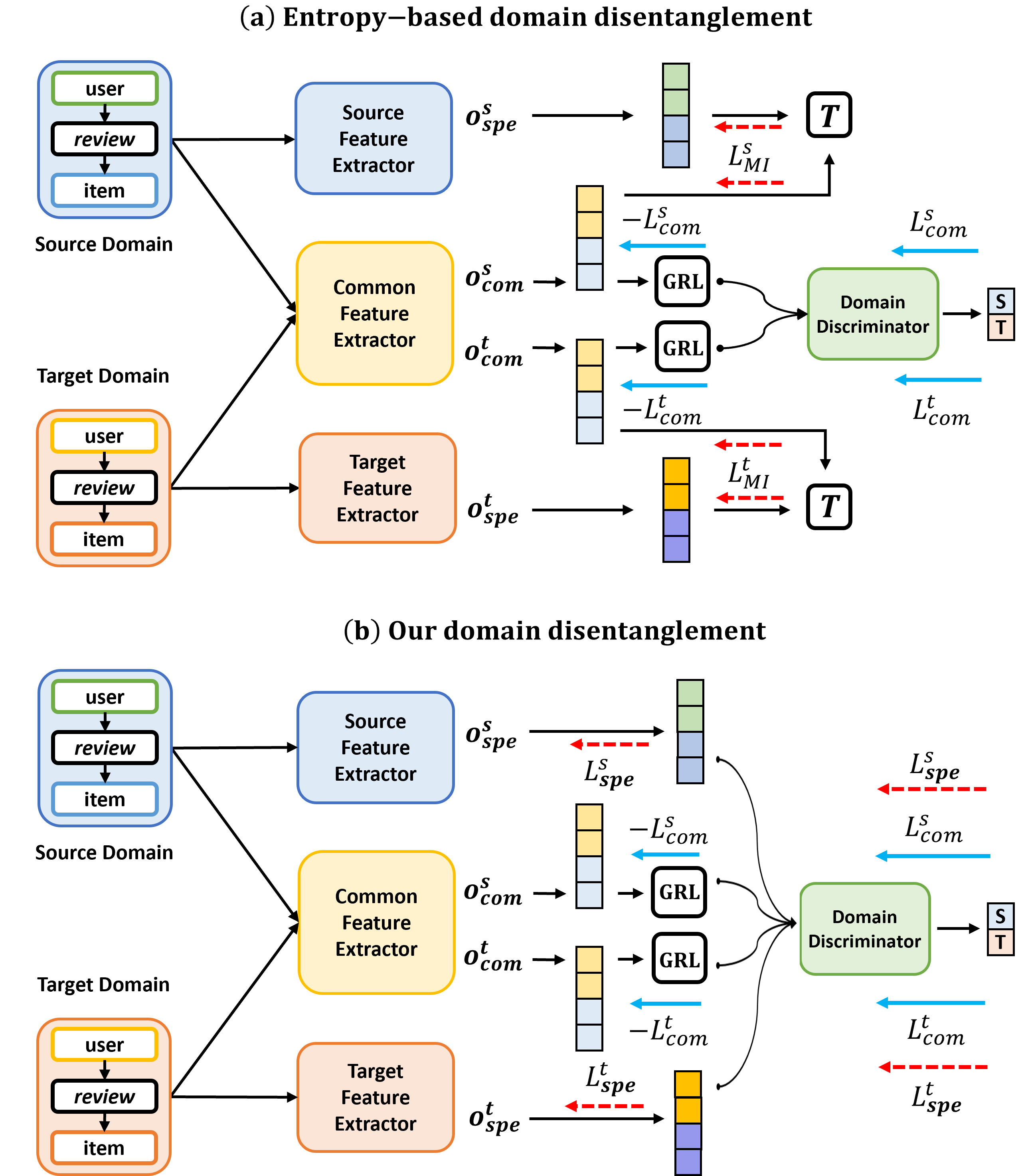}
  \caption{(a) Conventional domain disentanglement \cite{peng2019domain} using mutual information (MI) estimator $T$, and (b) ours integrating MI estimator with domain discriminator}
  \label{domain adaptation}
  \vspace{-3mm}
\end{figure}

\subsubsection{\textbf{Disentangling domain-specific features}}
To deal with the aforementioned problems, many approaches endeavored for the parallel extraction of domain-specific knowledge. One of the promising solutions is disentangled representation learning, which can secure pairwise independence between generated factors \cite{gretton2005measuring,li2019learning,nema2021disentangling}. 
For example, DSN \cite{bousmalis2016domain} optimizes the source, common, and target FEs by securing orthogonality between domain-specific and domain-common features. More recently, estimating mutual information (MI) between extracted features has been proposed \cite{peng2019domain,cheng2020improving}. Specifically, IDEL \cite{cheng2020improving} suggests a novel MI bound to retrieve disentangled representation of texts. DADA \cite{peng2019domain} captures domain-specific knowledge using MI estimator (see Figure \ref{domain adaptation}-a). In terms of independence maximization, DADA aims to minimize MI between domain-specific $O_{spe}$ and domain-common $O_{com}$ features as follows: 
\begin{equation}
\begin{gathered}
\mathcal{L}^s_{MI}= {1 \over N_s} \sum_{s=1}^{N_s} T(O^{s}_{spe},O^{s}_{com}) - log ({1 \over N_s} \sum_{s=1}^{N_s} e^{T(O^{s}_{spe},O'^{s}_{com})}), \\
\mathcal{L}^t_{MI}= {1 \over N_t} \sum_{t=1}^{N_t} T(O^{t}_{spe},O^{t}_{com}) - log ({1 \over N_t} \sum_{t=1}^{N_t} e^{T(O^{t}_{spe},O'^{t}_{com})}).
\end{gathered}
\label{mi_loss}
\end{equation}
Since MI cannot be directly calculated, they adopt approximation theory MINE \cite{belghazi2018mutual} with a neural network $T$: 
\begin{equation}
\begin{gathered}
\label{mine}
T(x, z) = F_3(\sigma(F_1(x) + F_2(z))).
\end{gathered}
\end{equation}
With two inputs $x$ and $z$, a fully-connected layer $F$ with non-linear activation $\sigma$ estimates MI. Back into Equation \ref{mi_loss}, as an input of $T$, ($O_{spe},O_{com}$) pairs are sampled from joint distribution, while the same number of ($O_{spe},O'_{com}$) pairs are from marginals (please refer to \cite{peng2019domain} for more details). 
As described in Figure \ref{domain adaptation}-a, domain-specific FEs are trained to minimize $\mathcal{L}^s_{MI},\mathcal{L}^t_{MI}$ (red arrows), while GRL guides a common FE (blue arrows) through Equation \ref{d_inv_loss}.
Nonetheless, MINE implicates the following limitations: a statistical drawback of measuring bounds \cite{mcallester2020formal}, the estimators become intractable even for a known distribution as the dimension increases \cite{gabrie2019entropy} and is highly dependent on the number of symbols and the size of dataset \cite{panzeri2007correcting}.  

Using the observations \cite{kang2019contrastive,saito2019semi,park2020joint,zhao2021domain}, minimizing $\mathcal{H}$-divergence is closely related to MI maximization. We integrate the MI estimator with a domain discriminator from the following perspectives: (1) maximizing domain loss (equal to MI maximization) for domain-common features, while (2) minimizing domain loss for domain-specific outputs. To achieve this, we exclude GRL between domain-specific FEs and domain discriminator (please refer to Figure \ref{domain adaptation}-b), to guide these FEs to generate domain-discriminative knowledge.
The advantages of this suggestion are as follows. Firstly, domain-specific features ($O^s_{spe}, O^t_{spe}$) are additionally utilized as an input of a domain discriminator, where they reinforce the discriminating power of a domain discriminator. Then, a discriminator well distinguishes the origin of the input vectors, leading three types of FEs better extract domain-aware knowledge. Secondly, our model can disentangle inter-domain outputs ($O^s_{spe} \leftrightarrow O^t_{spe}$), while the MI-based methods only separate intra-domain ones ($O^s_{spe} \leftrightarrow O^s_{com}$ or $O^t_{spe} \leftrightarrow O^t_{com}$).
Lastly, the large values of MI can mislead the parameters, while the ground truth of our model lies between 0 and 1. 
Using Equation \ref{domain_label} and \ref{d_inv_loss}, we can define domain-specific losses as:
\begin{equation} 
\begin{gathered}
\widehat{d}^s_{spe}=F_{disc}(O^s_{spe}), \, \,\widehat{d}^t_{spe}=F_{disc}(O^t_{spe}), \\
\mathcal{L}_{spe}^s= -{1 \over N_s}\sum_{s=1}^{N_s} log(1-\widehat{d}^s_{spe}),\,\,\mathcal{L}_{spe}^t = -{1 \over N_t}\sum_{t=1}^{N_t} log(\widehat{d}^t_{spe}).
\end{gathered}
\label{d_spe_loss}
\end{equation}

Finally, integrating Equation \ref{d_inv_loss} and \ref{d_spe_loss}, the overall domain loss function can be abbreviated as below:
\begin{equation}
\begin{gathered}
\mathcal{L}_{dom} = a(\mathcal{L}_{com}^s +
\mathcal{L}_{spe}^s) + (1-a)(\mathcal{L}_{com}^t +  \mathcal{L}_{spe}^t).
\end{gathered}
\label{da_loss}
\end{equation}
$a={N_s \over N_s+N_t}$ controls the weight of source and target losses based on the size of two domains \cite{ben2010theory}. For training, the domain discriminator is updated using Equation \ref{da_loss}. Then, GRL is applied for domain-common losses $-a\mathcal{L}^s_{com},-(1-a)\mathcal{L}^t_{com}$ (blue arrows in Figure \ref{domain adaptation}-b) to update common FE, while domain-specific losses $a\mathcal{L}^s_{spe},(1-a)\mathcal{L}^t_{spe}$ (red arrows) without GRL update domain-specific FEs. 

\subsubsection{Theoretical analysis}
Referring to the estimated bounds of semi-supervised DA \cite{li2021learning}, we can infer the theoretical bounds of DADA \cite{peng2019domain} and our SER. 
Since both of them captures domain-common knowledge through DA, the difference lies in domain disentanglement strategy and their performance. 
We first discuss the classification errors between true domain labels and predicted ones.
DADA adopts dual representations of KL-divergence, which has proven \cite{belghazi2018mutual} that a neural network $T_{\theta}$ with parameter $\theta$ satisfies the conditions below as the number of samples $n$ goes to infinity:
\begin{equation}
\begin{gathered}
|\widehat{I(X;Z)}_n-I_{\theta}(X,Z)| \leq \epsilon
\end{gathered}
\end{equation}
Similarly, our SER adopts stochastic gradient descent with binary domain labels, which has proven to \cite{gower2019sgd} converge under the arbitrary sampling scenario.
Based on these results, both approaches well predict the true labels within $\epsilon$. However, it is notable that the value of MI ranges from 0 to infinity. Referring to Equation \ref{mi_loss}, a large domain-specific loss might impede the stable convergence of three FEs. Comparatively, we adopt binary cross-entropy loss of Equation \ref{da_loss}, and thus, the loss rarely diverges even for completely domain-indistinguishable features ($\widehat{d}^s_{spe}\approx\widehat{d}^t_{spe}\approx0.5$). In Section \ref{dd_analysis}, we scrutinize two methods that use Equation \ref{mi_loss} and \ref{d_spe_loss}.

\subsection{\textbf{Encoding Network and Regressor}} 
\label{sec_re}
In this section, except for Equation \ref{enc_loss} and \ref{mse_loss}, we exclude domain identifier $d$ to simplify notation.

\textbf{Encoding network} This layer employs a transfer network \cite{wu2020learning} (MLP) to align the latent of aggregated reviews with an individual review, which is proposed by TransNet \cite{catherine2017transnets}. Using this strategy, FEs are trained to mimic the latent of an individual review that a user has written after purchasing a specific item (please refer to \cite{catherine2017transnets} for more details).
Here, we extend this mechanism from a single to cross-domain recommendation scenario, while most text-aided schemes \cite{zheng2017joint,chen2018neural,chen2019co,dong2020asymmetrical,wang2021leveraging} underestimate this kind of information.  
Since we have obtained the latents of aggregated review $O_{spe},O_{com}$, the encoded representation can be derived as below:
\begin{equation}
\label{E_out}
O  = F_{enc}(O_{spe}+O_{com})
\end{equation}
, where $F_{enc}$ is an encoding network with two layers.
Similarly, the representation of an individual review can be retrieved as follows:
\begin{equation}
\label{I_out}
I = I_{spe} + I_{com}.
\end{equation}

To align the two vectors $O\leftrightarrow I$ of Equation \ref{E_out} and \ref{I_out}, for each domain $d \in \{s,t\}$, we adopt Euclidean distance as a loss function: 
\begin{equation}
\label{enc_loss}
\mathcal{L}^d_{enc} = {1 \over N_d}\sum_{d=1}^{N_d}||O^d-I^d||_2^2.
\end{equation}

\textbf{Regressor} Finally, we can utilize two representations $O,I$ to predict a recommendation score through regressor $F_{reg}$ (MLP). Here, we further apply a widely used latent factor model \cite{tan2016rating}, where $p_u$ and $q_i$ stand for the embedding of user and item:
\begin{equation}
\label{regress_out}
\begin{split}
\widehat{y}_I=F_{reg}(I)+p_u\cdot q_i^T, \,\,\widehat{y}_O=F_{reg}(O)+p_u\cdot q_i^T.
\end{split}
\end{equation}
We utilize the prediction of an individual review $\widehat{y}_I$, which can guide the regressor precisely. The regression loss is defined in Equation \ref{mse_loss}, which is the Mean Squared Error (MSE) between two predicted scores $\widehat{y}_I$, $\widehat{y}_O$ and true label $y$ in domain $d$ as below:
\begin{equation}
\label{mse_loss}
\begin{split}
\mathcal{L}^d_{reg} = {1 \over 2N_d}\sum_{d = 1}^{N_d} \left((\widehat{y}^d_I-y^d)^2 + (\widehat{y}^d_O-y^d)^2\right).
\end{split}
\end{equation}

\subsection{Optimization and Inference}
\label{sec_opt}
\textbf{Optimization} We jointly optimize SER by minimizing the weighted sum of three losses defined in Equation \ref{da_loss}, \ref{enc_loss}, \ref{mse_loss}:
\begin{equation}
\label{total_loss}
\min_{\theta}\mathcal{L} = \alpha \mathcal{L}_{dom} + \beta (\mathcal{L}^s_{enc}+\mathcal{L}^t_{enc}) + \gamma (\mathcal{L}^s_{reg} + \mathcal{L}^t_{reg}) + \delta ||\theta||.
\end{equation}

In Figure \ref{model}, we describe how the gradients are propagated with three basic shapes (the horizontal stripe denotes a source loss). The hyper-parameters $\alpha$, $\beta$, and $\gamma$ balance the weight of three losses. We set $\alpha=0.1$, $\beta=0.05$, and $\gamma=1$ through a grid search. $\theta$ denotes parameters of our model, where $\delta$ is a regularization term. We adopt early stopping under 300 iterations with Adam optimizer and learning ratio as $lr=1e^{-4}$.

\textbf{Inference} During inference phase (lower side of Figure \ref{model}), SER recommends an item using aggregated reviews of user $R^t_u$ and item $R^t_i$.
Then, the common and target FEs are applied for $R^t_u$ and $R^t_i$, retrieving their features $O^t_{com},O^t_{spe}$, respectively. Finally, a target encoder generates meaningful feature $O^t$, followed by a regressor.

\subsection{Computational Complexity Analysis}
In this section, we analyze the computational cost of our model. Firstly, let us assume parameters of vanilla text analysis model DeepCoNN \cite{zheng2017joint} as $A+B$, which adopts a single FE ($A$) and regressor ($B$). The running time can be approximated as $\mathcal{O}((A+B) \cdot N_t \cdot e)$, where $N_t$ is the size of training samples and $e$ is a training epoch.

Our model further utilizes a domain discriminator with encoding networks. With the slight abuse of notation, let us assume the number of additional parameters as $C$. Also, we utilize the source domains that contain $N_s$ samples, the computational complexity can be abbreviated as $\mathcal{O}((3A+2B+C) \cdot (N_s+N_t) \cdot e)$. To summarize, we can infer that a cost linearly depends on the input size of the source domain $N_s$, which can be quite efficient.

\section{Experiments} \label{experiments}
We aim to answer the following research questions:
\begin{itemize}
    \item \textbf{RQ1:} Can SER enhance the recommendation quality compared to the state-of-the-art approaches? 
    \item \textbf{RQ2:} Does SER effectively disentangles features for a cross-domain recommendation?
    \item \textbf{RQ3:} How much does the domain-aware feature extractor and encoding network contribute to the overall quality? 
    \item \textbf{RQ4:} Is a domain-common feature necessary for CDR? If so, how does the knowledge is shared across domains?
\end{itemize}

\subsection{Dataset and Experimental Setup}
\textbf{Dataset description} We systematically conduct investigations with publicly available dataset \textit{Amazon}\footnote{http://jmcauley.ucsd.edu/data/amazon/} and \textit{Yelp}\footnote{https://www.yelp.com/dataset}. The target domain includes the following four categories of Amazon 5-core: \textit{Office Products, Instant Video, Automotive, Patio Lawn and Garden}. We designate the source domain with more interactions \cite{ben2010theory}. Three categories from \textit{Amazon}: \textit{Baby, Kindle Store (KS), Toys and Games (TG)} and one another from \textit{Yelp}. Here, \textit{Yelp} data is used to show the effect of excluding duplicate users. The statistical details of the dataset are summarized in Table \ref{dataset}.
Though some studies employ a source domain of seemingly relevant categories, it is natural to ask: \textit{how can we choose the most relevant domain? if they are truly relevant, how does the knowledge transferred across domains?} Here, we focus on the second question rather than figuring out relevant domains. Multi-source adaptation is also considerable, but here, we leave it for future work. 

\begin{table}
\caption{Statistical details of the dataset}
\label{dataset}
\centering
\begin{adjustbox}{width=0.48\textwidth}
\begin{tabular}{lllll}
\multicolumn{1}{l}{}    & \multicolumn{1}{l}{}       &         &         &                \\ 
\Xhline{2\arrayrulewidth}
\multicolumn{1}{l}{}        & Dataset                       & \# users & \# items & \# reviews  \\ 
\Xhline{2\arrayrulewidth}
\multirow{4}{*}{Source} 
                        & Baby                       & 19,445  & 7,050   & 160,792        \\
                        & Kindle Store (KS)               & 68,223  & 61,934  & 982,619        \\
                        & Toys and Games (TG)             & 19,412  & 11,924  & 167,597        \\
                        & Yelp             & 1.9 M  & 0.2 M  & 8.1 M        \\
\hline
\multirow{4}{*}{Target} & Office Products            & 4,905   & 2,420   & 53,258         \\
                        & Instant Video              & 5,130   & 1,685   & 37,126         \\
                        & Automotive                 & 2,928   & 1,835   & 20,473         \\
                        & Patio Lawn and Garden      & 1,686   & 962     & 13,272         \\
\Xhline{2\arrayrulewidth}
\end{tabular}
\end{adjustbox}
\end{table}

\begin{table*}[!htbp]
\centering
\caption{ MSE ($\downarrow$) on four target domain  dataset. Bold and underline indicate $1^{st}$ and $2^{nd}$ best.}
\label{performance_mse}
\small
\begin{tabularx}{\textwidth}{@{}c|cccc|cccc|cccc|cccc@{}}
\toprule
Target domain &  \multicolumn{4}{c|}{\underline{Office Product}} & \multicolumn{4}{c|}{\underline{Instant Video}} & \multicolumn{4}{c|}{\underline{Automotive}} & \multicolumn{4}{c}{\underline{Patio Lawn and Garden}} \\

Source domain & Baby & KS & TG & Yelp & Baby & KS & TG & Yelp & Baby & KS & TG & Yelp & Baby & KS & TG & Yelp \\
\midrule
PMF &  \multicolumn{4}{c|}{1.085} & \multicolumn{4}{c|}{1.129} & \multicolumn{4}{c|}{1.162} & \multicolumn{4}{c}{1.177} \\
NeuMF &  \multicolumn{4}{c|}{0.974} & \multicolumn{4}{c|}{1.014} & \multicolumn{4}{c|}{1.087} & \multicolumn{4}{c}{1.143} \\
DeepCoNN &  \multicolumn{4}{c|}{0.902} & \multicolumn{4}{c|}{0.949} & \multicolumn{4}{c|}{0.979} & \multicolumn{4}{c}{1.128} \\
NARRE &  \multicolumn{4}{c|}{0.863} & \multicolumn{4}{c|}{0.914} & \multicolumn{4}{c|}{0.887} & \multicolumn{4}{c}{1.108} \\
AHN &  \multicolumn{4}{c|}{0.859} & \multicolumn{4}{c|}{0.892} & \multicolumn{4}{c|}{0.863} & \multicolumn{4}{c}{\underline{1.094}} \\
\Xhline{2\arrayrulewidth}
DANN & 0.966 & 0.939 & 0.943 & 1.118 & 0.986 & 0.946 & 0.987 & 1.147 & 0.946 & 0.881 & 0.945 & 1.183 & 1.129 & 1.189 & 1.199 & 1.395\\
DAREC & 0.989 & 0.988 & 0.972 & 0.994 & 1.060 & 1.045 & 1.043 & 1.073 & 1.001 & 0.997 & 0.993 & 1.004 & 1.123 & 1.151 & 1.131 & 1.150\\
DDTCDR & 0.954 & 0.947 & 0.926 & 0.965 & 0.974 & 0.981 & 0.967 & 0.988 & 0.961 & 0.959 & 0.954 & 0.969 & 1.109 & 1.111 & 1.105 & 1.133\\ 
RC-DFM & 0.834 & 0.839 & 0.828 & \underline{0.841} & 0.878 & \underline{0.855} & \underline{0.868} & 0.872 & \underline{0.792} & \underline{0.800} & \underline{0.794} & \underline{0.802} & \underline{1.094} & 1.096 & 1.109 & 1.112\\
CATN & 0.875 & 0.872 & 0.873 & 0.876 & 0.915 & 0.906 & 0.892 & 0.919 & 0.824 & 0.831 & 0.826 & 0.837 & 1.141 & 1.144 & 1.129 & 1.149\\
MMT & \underline{0.815} & \underline{0.820} & \underline{0.822} & 0.856 & \underline{0.862} & 0.855 & 0.878 & \underline{0.871} & 0.818 & \textbf{0.798} & 0.800 & 0.833 & 1.116 & 1.099 & \underline{1.094} & 1.117\\
\Xhline{2\arrayrulewidth}
\textbf{SER} & \textbf{0.789} & \textbf{0.815} & \textbf{0.810} & \textbf{0.806} & \textbf{0.852} & \textbf{0.833} & \textbf{0.855} & \textbf{0.847} & \textbf{0.785} & \textbf{0.798} & \textbf{0.769} & \textbf{0.784} & \textbf{1.028} & \textbf{1.029} & \textbf{1.039} & \textbf{1.033}\\
\bottomrule
\end{tabularx}
\end{table*}

\begin{table*}[!htbp]
\centering
\small
\caption{ nDCG@5 ($\uparrow$) on four target domain dataset. Bold and underline indicate $1^{st}$ and $2^{nd}$ best. }
\label{performance_ndcg}
\begin{tabularx}{\textwidth}{@{}c|cccc|cccc|cccc|cccc@{}}
\toprule
Target domain &  \multicolumn{4}{c|}{\underline{Office Product}} & \multicolumn{4}{c|}{\underline{Instant Video}} & \multicolumn{4}{c|}{\underline{Automotive}} & \multicolumn{4}{c}{\underline{Patio Lawn and Garden}} \\
Source domain & Baby & KS & TG & Yelp & Baby & KS & TG & Yelp & Baby & KS & TG & Yelp & Baby & KS & TG & Yelp \\
\midrule
PMF &  \multicolumn{4}{c|}{0.737} & \multicolumn{4}{c|}{0.759} & \multicolumn{4}{c|}{0.764} & \multicolumn{4}{c}{0.770} \\
NeuMF &  \multicolumn{4}{c|}{0.756} & \multicolumn{4}{c|}{0.788} & \multicolumn{4}{c|}{0.781} & \multicolumn{4}{c}{0.776} \\
DeepCoNN &  \multicolumn{4}{c|}{0.856} & \multicolumn{4}{c|}{0.840} & \multicolumn{4}{c|}{0.816} & \multicolumn{4}{c}{0.842} \\
NARRE &  \multicolumn{4}{c|}{0.861} & \multicolumn{4}{c|}{0.872} & \multicolumn{4}{c|}{0.851} & \multicolumn{4}{c}{0.844} \\
AHN &  \multicolumn{4}{c|}{\underline{0.874}} & \multicolumn{4}{c|}{0.879} & \multicolumn{4}{c|}{0.862} & \multicolumn{4}{c}{\underline{0.878}} \\
\Xhline{2\arrayrulewidth}
DANN & 0.843 & 0.847 & 0.840  & 0.829  & 0.851 & 0.849  & 0.846  & 0.835  & 0.844 & 0.858  & 0.836  & 0.831  & 0.830 & 0.823  & 0.818  & 0.812 \\
DAREC & 0.859 & 0.842  & 0.835  & 0.827  & 0.844 & 0.848  & 0.841  & 0.823  & 0.865 & 0.872  & 0.872  & 0.860  & 0.854 & 0.852  & 0.861  & 0.835 \\
DDTCDR & 0.854 & 0.853 & 0.860 & 0.847 & 0.852 & 0.858 & 0.849 & 0.840 & 0.877 & 0.874 & 0.881 & 0.865 & 0.846 & 0.851 & 0.849 & 0.839\\
RC-DFM & 0.875  & 0.871  & \underline{0.880}  & 0.869  & \underline{0.890}  & 0.884  & \underline{0.881}  & 0.879  & 0.884  & 0.895  & \underline{0.899}  & \underline{0.902}  & \underline{0.878}  & 0.873  & 0.871  & \underline{0.879} \\
CATN & 0.869  & 0.865  & 0.871  & 0.842  &  0.873 & 0.857  & 0.860  & 0.873  & 0.866  & 0.863  & 0.872  & 0.875  & 0.864  & 0.861  & 0.858  & 0.854 \\
MMT & \underline{0.881} & \underline{0.874} & 0.870 & \underline{0.883}  & 0.888 & \underline{0.885} & 0.876 & \underline{0.883} & \underline{0.886} & \underline{0.896} & 0.892 & 0.877 & 0.867 & 0.869 & 0.871 & 0.871\\
\Xhline{2\arrayrulewidth}
\textbf{SER} & \textbf{0.891} & \textbf{0.885} & \textbf{0.888} & \textbf{0.889} & \textbf{0.896} & \textbf{0.892} & \textbf{0.889} & \textbf{0.895} & \textbf{0.892} & \textbf{0.901} & \textbf{0.908} & \textbf{0.913} & \textbf{0.889} & \textbf{0.882} & \textbf{0.885} & \textbf{0.883} \\
\bottomrule
\end{tabularx}
\end{table*}

\textbf{Environmental settings and baselines} Following previous studies, we divide the target domain dataset into three parts: 80 percent for training, 10 percent for validation, and another 10 percent for testing. We repeatedly consume a target domain w.r.t. the size of a source domain for one iteration. Early stopping is applied based on the validation score for 300 iterations. The word embedding is set to 100 with 100 convolution filters of size $\mathbb{R}^{5 \times 100}$. 
Now, we introduce state-of-the-art single and cross-domain methods below. 

\textit{Single-Domain Approaches:}
\begin{itemize}
    \item \textbf{PMF} \cite{mnih2008probabilistic} is a classical probabilistic matrix factorization method, which only utilizes rating information. 
    \item \textbf{NeuMF} \cite{he2017neural} combines deep neural networks with a probabilistic model, and use rating information only.
    \item \textbf{DeepCoNN} \cite{zheng2017joint} leverages review texts for rating prediction. They jointly encode the latent of user and item, respectively. 
    \item \textbf{NARRE} \cite{chen2018neural} maintains the overall architecture of \textit{DeepCoNN}, while employing attention mechanism. They firstly measure the usefulness of each review using attention scores.
    \item \textbf{AHN} \cite{dong2020asymmetrical} proposes a hierarchical attention mechanism: from a sentence to review-level attention for a better representation learning. It achieves state-of-the-art performance for a review-based single-domain recommendation.
\end{itemize}

\textit{Cross-Domain Approaches:}
\begin{itemize}
    \item \textbf{DANN} \cite{ganin2016domain} proposes the seminal domain adversarial technique that extracts domain-common features from two different domains. Here, review texts are embedded as 5,000-dimensional feature vectors.
    \item \textbf{DAREC} \cite{yuan2019darec} assume the same set of users between two domains. For each domain, the high-dimensional rating vectors are mapped to low-dimensional feature vectors using AutoEncoder, followed by a domain discriminator. They extract shareable rating patterns between two domains.
    \item \textbf{DDTCDR} \cite{li2020ddtcdr} assumes that if two users have similar preferences in a source domain, it should be preserved in a target domain through an orthogonal mapping function.  
    \item \textbf{RC-DFM} \cite{fu2019deeply} suggest fusing review texts with rating information. With SDAE, they adequately preserve the latent features with rich semantic information. Under our experimental setting, we fine-tune the text convolution layer. 
    \item \textbf{CATN} \cite{zhao2020catn} extracts multiple aspects from reviews. For a knowledge transfer between domains, they assume an aspect correlation matrix with an attention mechanism. 
    \item \textbf{MMT} \cite{krishnan2020transfer} suggest component-wise transfer mechanism, while also preserving domain-specific modules. Here, we assume the text convolution layer as a knowledge transfer module and fine-tune the parameters in a target domain.
\end{itemize}

\subsection{Results and Discussion (RQ1)}
For quantitative analysis, we assume two metrics: Mean Squared Error (MSE) for a rating prediction and normalized Discounted Cumulative Gain (nDCG$@5$) for ranking-based recommendations.

\textbf{Utilizing additional domain generally enhances recommendation quality} Table \ref{performance_mse}, \ref{performance_ndcg} shows the performance of our model and previous methods based on MSE and nDCG@5 score. For single-domain methodologies, it is not surprising that rating-based PMF and NeuMF performed worse than other review-based methods, which indicates the usefulness of textual information. Among review-based methods, NARRE and AHN outperform DeepCoNN with an attention mechanism. Nonetheless, it is noticeable that textual-based CDR (RC-DFM, CATN, MMT) generally outperforms single-domain methods under our experimental scenario.

\textbf{The quality of CDR can be degraded w.r.t. domain discrepancy} 
To test a domain discrepancy, we adopt DANN and MMT, where they transfer knowledge independent of user overlapping. For DANN, the result varies significantly w.r.t. the selection of source domain. Rather, MMT secures stability through domain-specific information (rating).
An important thing is that a CDR does not always achieve the best performance.
In the case of \textit{Patio Lawn and Garden}, AHN shows the best result among all baselines. It indicates that utilizing additional domains without debilitating noises can degrade the recommendation quality. Comparatively, SER manages to achieve stable performance, where the domain disentanglement alleviates noises of the auxiliary domain. 

\textbf{Instead of overlapping users, review-based knowledge transfer further improves the recommendation quality} 
For example, RC-DFM integrates textual data with rating information, which significantly outperforms DAREC and DDTCDR. 
However, the problem is that all these methods require duplicate users for a knowledge transfer. This can be quite restrictive for the selection of a source domain, and some argued \cite{kang2019semi} that user-based knowledge transfer has a limited impact. Even excluding duplicate users (selecting $Yelp$), their performance varies quite insignificant (please refer to \cite{fu2019deeply} for more details) compared to the other methods (CATN, MMT, SER). Comparatively, MMT and SER (review-based transfer) improve their score even by selecting \textit{Yelp} as a source, which fairly is different from \textit{Amazon} dataset.
Though CATN utilizes auxiliary reviews of another user, does not show outstanding performance in our experiments. Further, MMT shows lower performance than SER because of the limitation of transfer learning and the absence of review disentanglement.
Using the above results, we show that SER surpasses the SOTA methods without duplicate users or contexts.

\subsection{\textbf{Analysis of Domain Discriminator (RQ2)}} \label{dd_analysis}
In this section, we compare state-of-the-art the domain disentanglement strategy \cite{peng2019domain} with our model, while numerous optimization methods can be considered \cite{gretton2005measuring,li2019learning,bousmalis2016domain}. We assume two different models; (1) $SER_{MI}$ minimizes mutual information \cite{peng2019domain}, while other modules remain the same as $SER$, and (2) our original $SER_{SA}$. To implement $SER_{MI}$, we utilize the publicly available source \textit{code}\footnote{https://github.com/VisionLearningGroup/DAL}. Specifically, we train them to minimize the absolute value of estimated MI (please refer to Equation \ref{mi_loss}).

\begin{figure}
  \vspace{-2mm}
\includegraphics[width=0.49\textwidth]{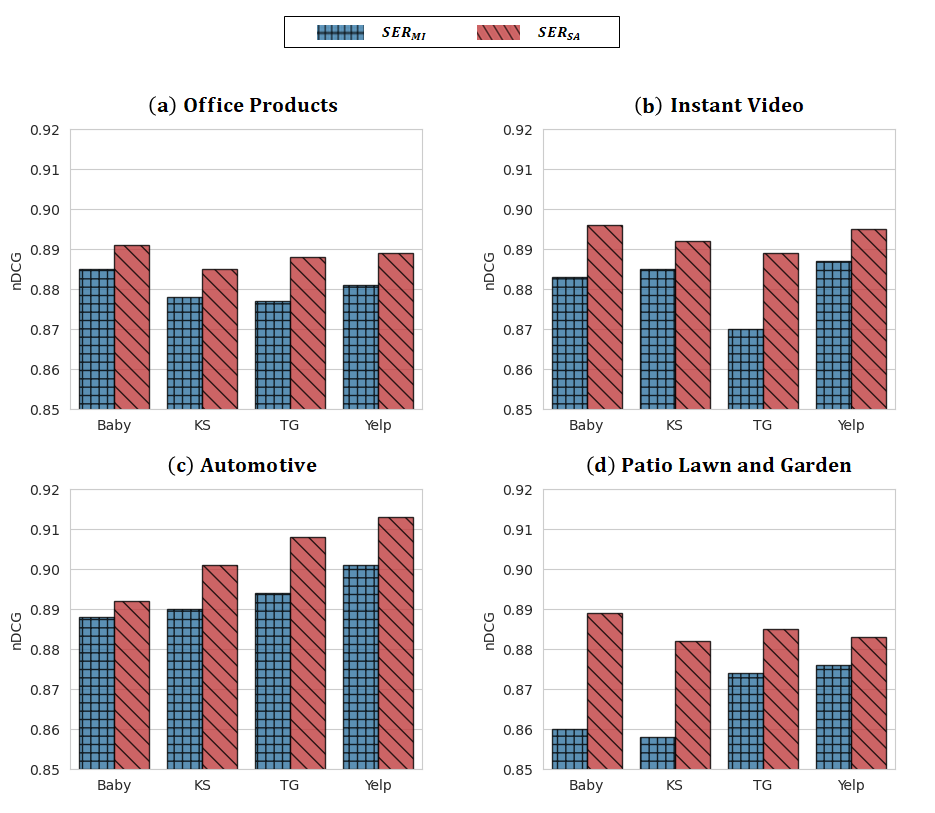}
  \caption{To show the influence of domain loss function, we describe nDCG@5 ($\uparrow$) of two SER variants; SER with MI minimization $SER_{MI}$, and our proposed model $SER_{SA}$}
  \label{da.per}
\end{figure}

\begin{figure}
  \hspace{-3mm}
 \includegraphics[width=0.48\textwidth]{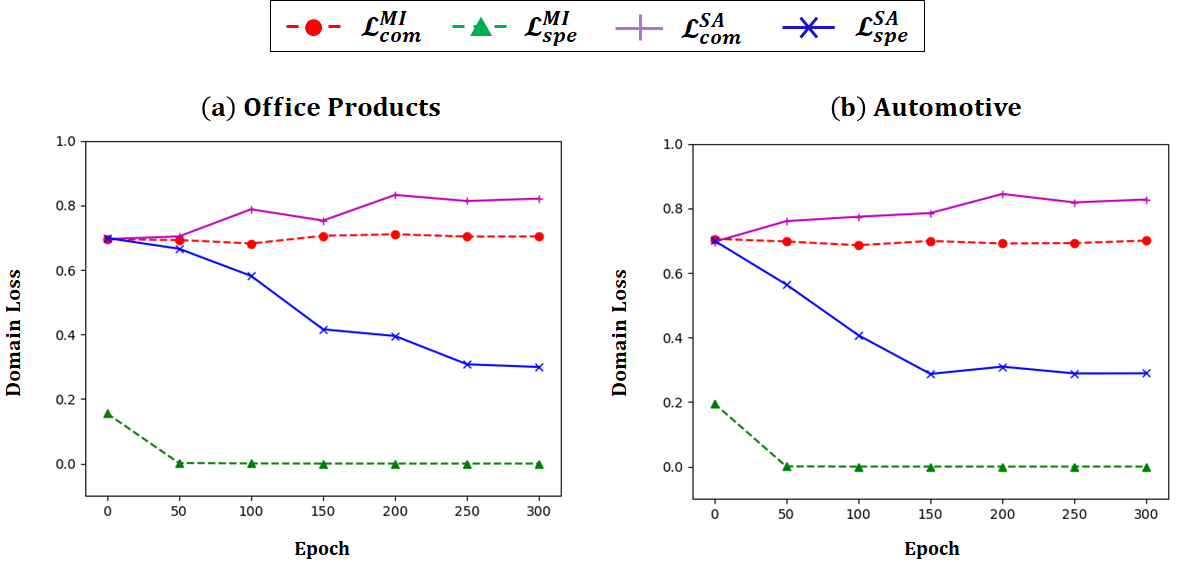}
  \caption{For two SER variants; $SER_{MI}$ and $SER_{SA}$, we plot their domain losses \textit{w.r.t.} training epoch}
  \label{da.loss}
\end{figure}

\textbf{Analysis of the recommendation quality \textit{w.r.t.} domain loss} To scrutinize how the domain loss affects the overall performance, we measure the nDCG@5 of two variants using four target domain data, which is illustrated in Figure \ref{da.per}.
Here, we notice that our $SER_{SA}$ slightly outperforms $SER_{MI}$ for all CDR scenarios. We insist that these results are closely related to domain losses. In Figure \ref{da.loss}, using $Baby$ as a source with two target domain datasets (\textit{Office Products, Automotive}), we plot the variation of domain losses for $SER_{MI}$ and $SER_{SA}$ \textit{w.r.t.} training epoch. Specifically, (1) for domain-common loss, both of them adopts adversarial training of Equation \ref{d_inv_loss}, $\mathcal{L}_{com}^{MI}=\mathcal{L}_{com}^{SA}$=$\mathcal{L}_{com}^s+\mathcal{L}_{com}^t$. (2) instead, their domain-specific loss ($\mathcal{L}_{spe}^{MI}$, $\mathcal{L}_{spe}^{SA}$) differs. $SER_{MI}$ adopts Equation \ref{mi_loss}, while $SER_{SA}$ utilizes Equation \ref{d_spe_loss}.
Referring Figure \ref{da.loss}, for $SER_{MI}$, $\mathcal{L}_{com}^{MI}$ (red line) rarely changes, while $\mathcal{L}_{spe}^{MI}$ (green) converges so fast. Comparatively, $\mathcal{L}_{com}^{SA}$ and $\mathcal{L}_{spe}^{SA}$ slowly converges, consistently pushing three FEs for domain disentanglement.
This result indicates that integrating the MI estimator with a domain discriminator not only secures robustness but also achieves better performance.

\begin{figure}
  \hspace{-2mm}
  \includegraphics[width=0.49\textwidth]{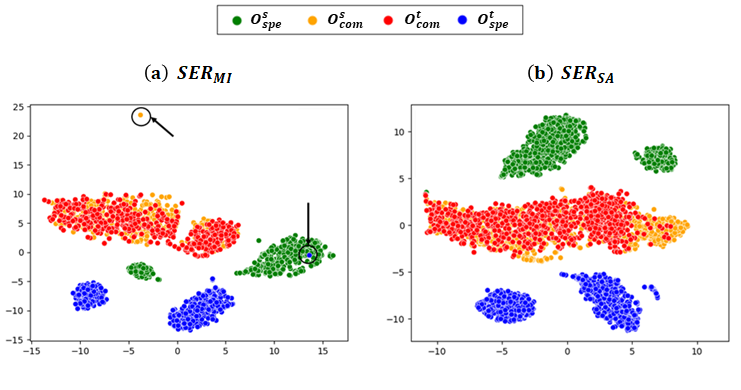}
  \caption{Visualization of disentangled representations for two SER variants; $SER_{MI}$ and $SER_{SA}$. From the source and target domains, we randomly sample user and item pairs, and plotted their reviews using t-SNE}
  \label{da.viz}
  \vspace{-3mm}
\end{figure}

\textbf{Visualization of disentangled representations}
In Figure \ref{da.viz}, we visualize the disentangled representations of two \textit{SER} variants. For an experiment, we pre-trained our model using \textit{Baby} and \textit{Patio Lawn and Garden} as source and target domain data.
Here, we randomly select one thousand user-item pairs, and extract four types of features $O^s_{spe},O^s_{com},O^t_{com},O^t_{spe}$ from their aggregated reviews. The visualization is conducted using t-distributed Stochastic Neighbor Embedding (t-SNE), which is called a nonlinear dimensionality reduction method for high dimensional representations. In detail, t-SNE has shown to reflect the similarity between data points, where we set the perplexity as 30.

Through Figure \ref{da.viz}-a and b, we can see that both methods well capture domain-common knowledge ($O^s_{com},O^t_{com}$ are overlapped each other). Further, two methods seem to be well disentangle domain-specific features ($O^s_{spe},O^t_{spe}$) from domain-common outputs.
However, for a disentanglement between domain-specific features ($O^s_{spe}$ $\leftrightarrow$ $O^t_{spe}$), $SER_{MI}$ has shown to implicate some defects. Firstly, in Figure \ref{da.viz}-a, $O^s_{spe}$ (cluster of green points) are positioned between $O^t_{spe}$ (cluster of blue points). As mentioned above, t-SNE preserves a point-wise distance after dimensionality reduction, but the clusters derived from $SER_{MI}$ fall apart from each other. Secondly, some points appear out of their original cluster (marked with a circle). We argue that the objective function of $SER_{MI}$ only minimizes MI of each domain independently while disregarding discrimination between domain-specific features. 
Comparatively, in Figure \ref{da.viz}-b, we notice that $SER_{SA}$ well captures domain-specific knowledge, where the outputs are included in a dense region without outliers.

\subsection{Ablation Study (RQ3)}
\begin{figure}
\hspace{-3mm}
    \includegraphics[width=0.49\textwidth]{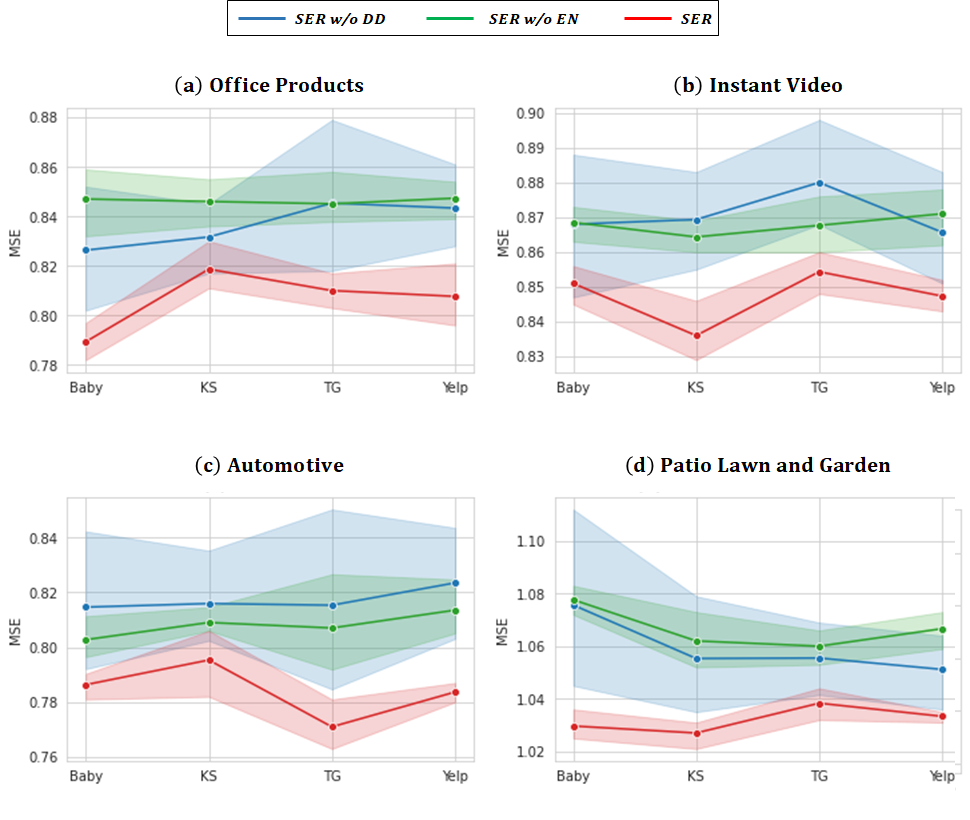}
  \caption{Ablation study (RQ3) of three SER variants. We adopt MSE ($\downarrow$) score for a comparison}
  \label{Fig.module}
\end{figure}

To test the significance of domain-aware feature extractor and encoding network, we assume three variants: (1) \textit{SER w/o DD}: excluding domain discriminator, (2) \textit{SER w/o EN}: excluding encoding network, and (3) our original \textit{SER}. Like previous studies, we conduct experiments under four target domain data.
In Figure \ref{Fig.module}, for each domain, we plot the min, median, and max value of MSE scores under 5 iterations. We do not assume a model excluding both components, which is identical to the vanilla text analysis model \cite{zheng2017joint}.

\textbf{Domain-aware feature extractor is fundamental for debilitating noises} We discuss \textit{SER w/o DD} (blue lines), removing domain discriminator from SER. In Figure \ref{Fig.module}, we can see that the performance of \textit{SER w/o DD} shows higher variance compared to \textit{SER w/o EN} (green lines), suggesting the domain discrepancy is critical for the overall performance in a target domain (the column-wise width denotes the variance of the results). For example, in Figure \ref{Fig.module}-b (\textit{Instant Video}), we notice that the adoption of \textit{Yelp} as a source domain improves the performance, whereas utilizing \textit{Toys and Games} can degrade the recommendation quality.

\textbf{Encoding network contributes to the recommendation quality significantly} Compared to \textit{SER w/o DD}, \textit{SER w/o EN} shows relatively stable results, which can debilitate noises from a source domain. Though the domain-aware feature extractor effectively controls the mismatches between two different domains, we notice that excluding the encoding layer is critical for the overall performance. Specifically, for all datasets, it is notable that \textit{SER w/o DD} slightly outperforms \textit{SER w/o EN}.

\subsection{Case Study (RQ4)}
\begin{figure}
  \includegraphics[width=0.48\textwidth]{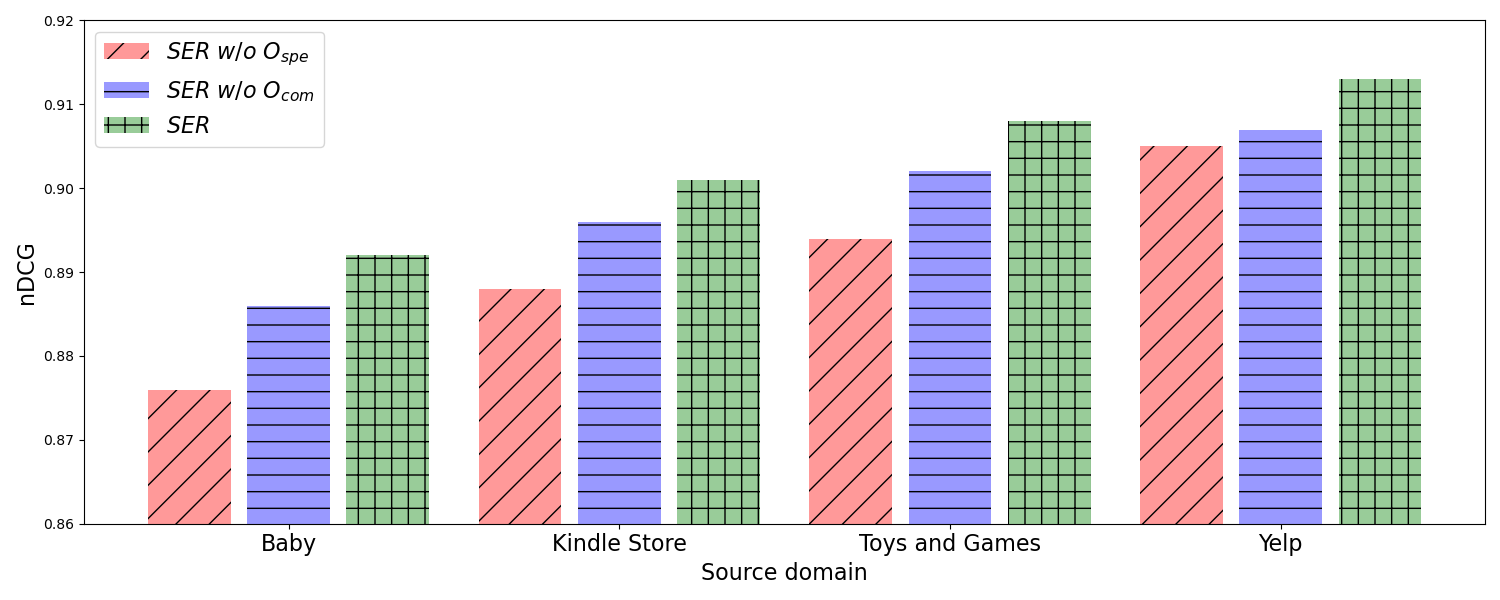}
  \caption{Case study (RQ4) of three SER variants. We adopt nDCG@5 ($\uparrow$) score for a comparison} 
  \label{case_study}
\end{figure}

In this section, we scrutinize the interpretability of SER through a real-world dataset $Amazon$. In Figure \ref{case_study}, using \textit{Automotive} as a target data, we describe the nDCG@5 of three methods; SER w/o domain-specific features $O_{spe}$, SER w/o domain-common features $O_{com}$, and our original $SER$. As can be seen, excluding a single type of information downgrades the performance. Here, we notice that domain-specific information contributes slightly more to the overall quality since $SER$ $w/o$ $O_{com}$ outperforms $SER$ $w/o$ $O_{spe}$ $1.2\%$ on average. We now scrutinize this phenomenon below.

\begin{figure}
  \includegraphics[width=0.48\textwidth]{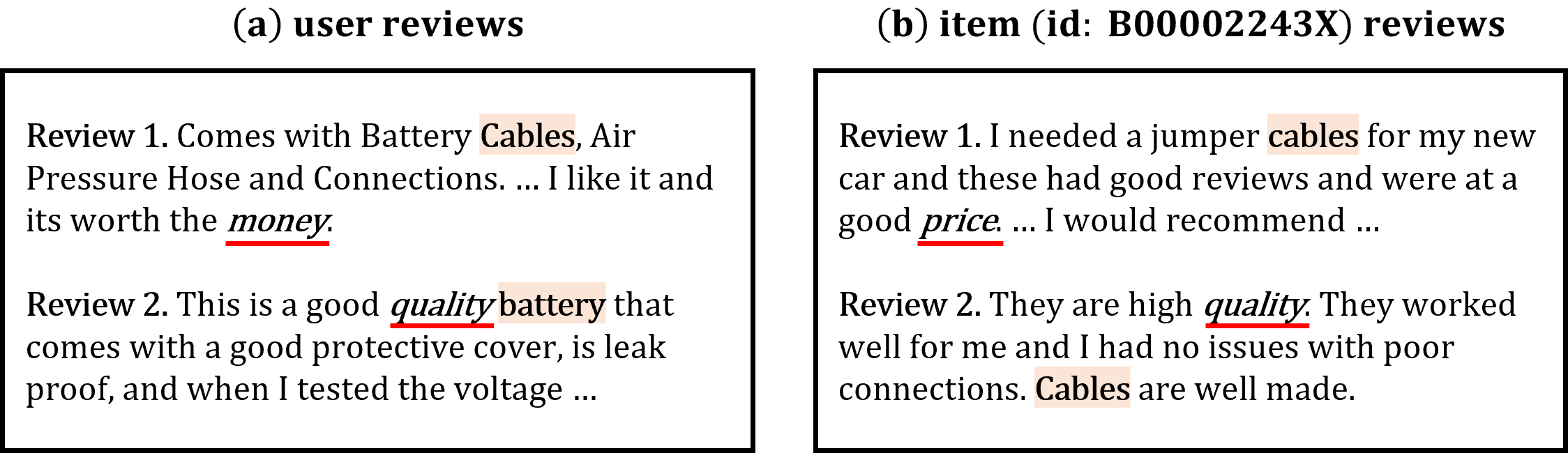}
  \caption{Explainability analysis for user and item reviews}
  \label{eval_ex}
  \vspace{-3mm}
\end{figure}

In Figure \ref{eval_ex}, we pre-trained a model using $Baby$ for $Automotive$. Then, we apply common and target FEs for each review, retrieving domain-common $O_{com}$, and domain-specific $O_{spe}$ outputs. We \textit{italicized} the most similar words (\textit{e.g., cosine similarity}) of domain-common feature $O_{com}$, and colorized most similar words of domain-specific ones $O_{spe}$. As can be seen, the italicized words further contain semantic meaning (\textit{quality, price, and money}), while the colorized one contains the user's own preference (\textit{cable and battery}). Throughout these results, we can infer that SER well disentangles domain knowledge since two features are complimentary, while none of them are negligible for the improvement of overall quality.

\section{Conclusion}
In this paper, we propose a novel method of utilizing review texts in multiple domains without requiring overlapping users or contexts. Our optimization strategies for domain disentanglement well achieve the knowledge transfer between domains, while also securing robustness. Further, we successfully extend a transfer network from a single to multiple domains, which can infer the latent of an individual review regarding domain-aware knowledge. The extensive experiments and ablation studies demonstrate the superiority and advantages of our method for the CDR scenario.

\subsubsection*{\normalfont{\textbf{Acknowledgments}}}
This work was supported by the National Research Foundation of Korea (NRF) (No. 2016R1A5A1012966, No. 2020R1A2C110168713),  Institute of Information \& communications Technology Planning \& Evaluation (IITP) (No. 2021-0-02068 Artificial Intelligence Innovation Hub, No. RS-2022-00156287 Innovative Human Resource Development for Local Intellectualization support program) grant funded by the Korea government (MSIT).


\bibliographystyle{ACM-Reference-Format}
\balance
\bibliography{references.bib}


\begin{thebibliography}{67}


\ifx \showCODEN    \undefined \def \showCODEN     #1{\unskip}     \fi
\ifx \showDOI      \undefined \def \showDOI       #1{#1}\fi
\ifx \showISBNx    \undefined \def \showISBNx     #1{\unskip}     \fi
\ifx \showISBNxiii \undefined \def \showISBNxiii  #1{\unskip}     \fi
\ifx \showISSN     \undefined \def \showISSN      #1{\unskip}     \fi
\ifx \showLCCN     \undefined \def \showLCCN      #1{\unskip}     \fi
\ifx \shownote     \undefined \def \shownote      #1{#1}          \fi
\ifx \showarticletitle \undefined \def \showarticletitle #1{#1}   \fi
\ifx \showURL      \undefined \def \showURL       {\relax}        \fi
\providecommand\bibfield[2]{#2}
\providecommand\bibinfo[2]{#2}
\providecommand\natexlab[1]{#1}
\providecommand\showeprint[2][]{arXiv:#2}

\bibitem[\protect\citeauthoryear{Banerjee, Bhattacharyya, and Bose}{Banerjee
  et~al\mbox{.}}{2017}]%
        {banerjee2017whose}
\bibfield{author}{\bibinfo{person}{Shankhadeep Banerjee},
  \bibinfo{person}{Samadrita Bhattacharyya}, {and} \bibinfo{person}{Indranil
  Bose}.} \bibinfo{year}{2017}\natexlab{}.
\newblock \showarticletitle{Whose online reviews to trust? Understanding
  reviewer trustworthiness and its impact on business}.
\newblock \bibinfo{journal}{\emph{Decision Support Systems}}
  \bibinfo{volume}{96} (\bibinfo{year}{2017}), \bibinfo{pages}{17--26}.
\newblock


\bibitem[\protect\citeauthoryear{Bao, Fang, and Zhang}{Bao
  et~al\mbox{.}}{2014}]%
        {bao2014topicmf}
\bibfield{author}{\bibinfo{person}{Yang Bao}, \bibinfo{person}{Hui Fang}, {and}
  \bibinfo{person}{Jie Zhang}.} \bibinfo{year}{2014}\natexlab{}.
\newblock \showarticletitle{Topicmf: Simultaneously exploiting ratings and
  reviews for recommendation}. In \bibinfo{booktitle}{\emph{Twenty-Eighth AAAI
  conference on artificial intelligence}}.
\newblock


\bibitem[\protect\citeauthoryear{Belghazi, Baratin, Rajeshwar, Ozair, Bengio,
  Courville, and Hjelm}{Belghazi et~al\mbox{.}}{2018}]%
        {belghazi2018mutual}
\bibfield{author}{\bibinfo{person}{Mohamed~Ishmael Belghazi},
  \bibinfo{person}{Aristide Baratin}, \bibinfo{person}{Sai Rajeshwar},
  \bibinfo{person}{Sherjil Ozair}, \bibinfo{person}{Yoshua Bengio},
  \bibinfo{person}{Aaron Courville}, {and} \bibinfo{person}{Devon Hjelm}.}
  \bibinfo{year}{2018}\natexlab{}.
\newblock \showarticletitle{Mutual information neural estimation}. In
  \bibinfo{booktitle}{\emph{International Conference on Machine Learning}}.
  PMLR, \bibinfo{pages}{531--540}.
\newblock


\bibitem[\protect\citeauthoryear{Ben-David, Blitzer, Crammer, Kulesza, Pereira,
  and Vaughan}{Ben-David et~al\mbox{.}}{2010}]%
        {ben2010theory}
\bibfield{author}{\bibinfo{person}{Shai Ben-David}, \bibinfo{person}{John
  Blitzer}, \bibinfo{person}{Koby Crammer}, \bibinfo{person}{Alex Kulesza},
  \bibinfo{person}{Fernando Pereira}, {and} \bibinfo{person}{Jennifer~Wortman
  Vaughan}.} \bibinfo{year}{2010}\natexlab{}.
\newblock \showarticletitle{A theory of learning from different domains}.
\newblock \bibinfo{journal}{\emph{Machine learning}} \bibinfo{volume}{79},
  \bibinfo{number}{1} (\bibinfo{year}{2010}), \bibinfo{pages}{151--175}.
\newblock


\bibitem[\protect\citeauthoryear{Bonab, Aliannejadi, Vardasbi, Kanoulas, and
  Allan}{Bonab et~al\mbox{.}}{2021}]%
        {bonab2021cross}
\bibfield{author}{\bibinfo{person}{Hamed Bonab}, \bibinfo{person}{Mohammad
  Aliannejadi}, \bibinfo{person}{Ali Vardasbi}, \bibinfo{person}{Evangelos
  Kanoulas}, {and} \bibinfo{person}{James Allan}.}
  \bibinfo{year}{2021}\natexlab{}.
\newblock \showarticletitle{Cross-Market Product Recommendation}. In
  \bibinfo{booktitle}{\emph{Proceedings of the 30th ACM International
  Conference on Information \& Knowledge Management}}.
  \bibinfo{pages}{110--119}.
\newblock


\bibitem[\protect\citeauthoryear{Bousmalis, Trigeorgis, Silberman, Krishnan,
  and Erhan}{Bousmalis et~al\mbox{.}}{2016}]%
        {bousmalis2016domain}
\bibfield{author}{\bibinfo{person}{Konstantinos Bousmalis},
  \bibinfo{person}{George Trigeorgis}, \bibinfo{person}{Nathan Silberman},
  \bibinfo{person}{Dilip Krishnan}, {and} \bibinfo{person}{Dumitru Erhan}.}
  \bibinfo{year}{2016}\natexlab{}.
\newblock \showarticletitle{Domain separation networks}.
\newblock \bibinfo{journal}{\emph{Advances in neural information processing
  systems}}  \bibinfo{volume}{29} (\bibinfo{year}{2016}),
  \bibinfo{pages}{343--351}.
\newblock


\bibitem[\protect\citeauthoryear{Cai, Li, Wei, Qiao, Zhang, and Hao}{Cai
  et~al\mbox{.}}{2019}]%
        {cai2019learning}
\bibfield{author}{\bibinfo{person}{Ruichu Cai}, \bibinfo{person}{Zijian Li},
  \bibinfo{person}{Pengfei Wei}, \bibinfo{person}{Jie Qiao},
  \bibinfo{person}{Kun Zhang}, {and} \bibinfo{person}{Zhifeng Hao}.}
  \bibinfo{year}{2019}\natexlab{}.
\newblock \showarticletitle{Learning disentangled semantic representation for
  domain adaptation}. In \bibinfo{booktitle}{\emph{IJCAI: proceedings of the
  conference}}, Vol.~\bibinfo{volume}{2019}. NIH Public Access,
  \bibinfo{pages}{2060}.
\newblock


\bibitem[\protect\citeauthoryear{Catherine and Cohen}{Catherine and
  Cohen}{2017}]%
        {catherine2017transnets}
\bibfield{author}{\bibinfo{person}{Rose Catherine} {and}
  \bibinfo{person}{William Cohen}.} \bibinfo{year}{2017}\natexlab{}.
\newblock \showarticletitle{Transnets: Learning to transform for
  recommendation}. In \bibinfo{booktitle}{\emph{Proceedings of the eleventh ACM
  conference on recommender systems}}. \bibinfo{pages}{288--296}.
\newblock


\bibitem[\protect\citeauthoryear{Chen, Zhang, Liu, and Ma}{Chen
  et~al\mbox{.}}{2018}]%
        {chen2018neural}
\bibfield{author}{\bibinfo{person}{Chong Chen}, \bibinfo{person}{Min Zhang},
  \bibinfo{person}{Yiqun Liu}, {and} \bibinfo{person}{Shaoping Ma}.}
  \bibinfo{year}{2018}\natexlab{}.
\newblock \showarticletitle{Neural attentional rating regression with
  review-level explanations}. In \bibinfo{booktitle}{\emph{Proceedings of the
  2018 World Wide Web Conference}}. \bibinfo{pages}{1583--1592}.
\newblock


\bibitem[\protect\citeauthoryear{Chen, Wang, Fu, Long, and Wang}{Chen
  et~al\mbox{.}}{2019a}]%
        {chen2019catastrophic}
\bibfield{author}{\bibinfo{person}{Xinyang Chen}, \bibinfo{person}{Sinan Wang},
  \bibinfo{person}{Bo Fu}, \bibinfo{person}{Mingsheng Long}, {and}
  \bibinfo{person}{Jianmin Wang}.} \bibinfo{year}{2019}\natexlab{a}.
\newblock \showarticletitle{Catastrophic forgetting meets negative transfer:
  Batch spectral shrinkage for safe transfer learning}.
\newblock  (\bibinfo{year}{2019}).
\newblock


\bibitem[\protect\citeauthoryear{Chen, Wang, Long, and Wang}{Chen
  et~al\mbox{.}}{2019b}]%
        {chen2019transferability}
\bibfield{author}{\bibinfo{person}{Xinyang Chen}, \bibinfo{person}{Sinan Wang},
  \bibinfo{person}{Mingsheng Long}, {and} \bibinfo{person}{Jianmin Wang}.}
  \bibinfo{year}{2019}\natexlab{b}.
\newblock \showarticletitle{Transferability vs. discriminability: Batch
  spectral penalization for adversarial domain adaptation}. In
  \bibinfo{booktitle}{\emph{International conference on machine learning}}.
  PMLR, \bibinfo{pages}{1081--1090}.
\newblock


\bibitem[\protect\citeauthoryear{Chen, Zhang, and Qin}{Chen
  et~al\mbox{.}}{2019d}]%
        {chen2019dynamic}
\bibfield{author}{\bibinfo{person}{Xu Chen}, \bibinfo{person}{Yongfeng Zhang},
  {and} \bibinfo{person}{Zheng Qin}.} \bibinfo{year}{2019}\natexlab{d}.
\newblock \showarticletitle{Dynamic Explainable Recommendation based on Neural
  Attentive Models}. In \bibinfo{booktitle}{\emph{Proceedings of the AAAI
  Conference on Artificial Intelligence}}, Vol.~\bibinfo{volume}{33}.
  \bibinfo{pages}{53--60}.
\newblock


\bibitem[\protect\citeauthoryear{Chen, Wang, Xie, Wu, Bu, Wang, and Chen}{Chen
  et~al\mbox{.}}{2019c}]%
        {chen2019co}
\bibfield{author}{\bibinfo{person}{Zhongxia Chen}, \bibinfo{person}{Xiting
  Wang}, \bibinfo{person}{Xing Xie}, \bibinfo{person}{Tong Wu},
  \bibinfo{person}{Guoqing Bu}, \bibinfo{person}{Yining Wang}, {and}
  \bibinfo{person}{Enhong Chen}.} \bibinfo{year}{2019}\natexlab{c}.
\newblock \showarticletitle{Co-attentive multi-task learning for explainable
  recommendation}. In \bibinfo{booktitle}{\emph{Proceedings of the 28th
  International Joint Conference on Artificial Intelligence}}. AAAI Press,
  \bibinfo{pages}{2137--2143}.
\newblock


\bibitem[\protect\citeauthoryear{Cheng, Min, Shen, Malon, Zhang, Li, and
  Carin}{Cheng et~al\mbox{.}}{2020}]%
        {cheng2020improving}
\bibfield{author}{\bibinfo{person}{Pengyu Cheng},
  \bibinfo{person}{Martin~Renqiang Min}, \bibinfo{person}{Dinghan Shen},
  \bibinfo{person}{Christopher Malon}, \bibinfo{person}{Yizhe Zhang},
  \bibinfo{person}{Yitong Li}, {and} \bibinfo{person}{Lawrence Carin}.}
  \bibinfo{year}{2020}\natexlab{}.
\newblock \showarticletitle{Improving disentangled text representation learning
  with information-theoretic guidance}.
\newblock \bibinfo{journal}{\emph{arXiv preprint arXiv:2006.00693}}
  (\bibinfo{year}{2020}).
\newblock


\bibitem[\protect\citeauthoryear{Dong, Ni, Cheng, Chen, Zong, Song, Liu, Chen,
  and De~Melo}{Dong et~al\mbox{.}}{2020}]%
        {dong2020asymmetrical}
\bibfield{author}{\bibinfo{person}{Xin Dong}, \bibinfo{person}{Jingchao Ni},
  \bibinfo{person}{Wei Cheng}, \bibinfo{person}{Zhengzhang Chen},
  \bibinfo{person}{Bo Zong}, \bibinfo{person}{Dongjin Song},
  \bibinfo{person}{Yanchi Liu}, \bibinfo{person}{Haifeng Chen}, {and}
  \bibinfo{person}{Gerard De~Melo}.} \bibinfo{year}{2020}\natexlab{}.
\newblock \showarticletitle{Asymmetrical hierarchical networks with attentive
  interactions for interpretable review-based recommendation}. In
  \bibinfo{booktitle}{\emph{Proceedings of the AAAI Conference on Artificial
  Intelligence}}, Vol.~\bibinfo{volume}{34}. \bibinfo{pages}{7667--7674}.
\newblock


\bibitem[\protect\citeauthoryear{Elkahky, Song, and He}{Elkahky
  et~al\mbox{.}}{2015}]%
        {elkahky2015multi}
\bibfield{author}{\bibinfo{person}{Ali~Mamdouh Elkahky}, \bibinfo{person}{Yang
  Song}, {and} \bibinfo{person}{Xiaodong He}.} \bibinfo{year}{2015}\natexlab{}.
\newblock \showarticletitle{A multi-view deep learning approach for cross
  domain user modeling in recommendation systems}. In
  \bibinfo{booktitle}{\emph{Proceedings of the 24th International Conference on
  World Wide Web}}. \bibinfo{pages}{278--288}.
\newblock


\bibitem[\protect\citeauthoryear{Fu, Zhang, Hu, Dai, Huang, and Chen}{Fu
  et~al\mbox{.}}{2021}]%
        {fu2021dual}
\bibfield{author}{\bibinfo{person}{Bairan Fu}, \bibinfo{person}{Wenming Zhang},
  \bibinfo{person}{Guangneng Hu}, \bibinfo{person}{Xinyu Dai},
  \bibinfo{person}{Shujian Huang}, {and} \bibinfo{person}{Jiajun Chen}.}
  \bibinfo{year}{2021}\natexlab{}.
\newblock \showarticletitle{Dual Side Deep Context-aware Modulation for Social
  Recommendation}. In \bibinfo{booktitle}{\emph{Proceedings of the Web
  Conference 2021}}. \bibinfo{pages}{2524--2534}.
\newblock


\bibitem[\protect\citeauthoryear{Fu, Peng, Wang, Xu, and Li}{Fu
  et~al\mbox{.}}{2019}]%
        {fu2019deeply}
\bibfield{author}{\bibinfo{person}{Wenjing Fu}, \bibinfo{person}{Zhaohui Peng},
  \bibinfo{person}{Senzhang Wang}, \bibinfo{person}{Yang Xu}, {and}
  \bibinfo{person}{Jin Li}.} \bibinfo{year}{2019}\natexlab{}.
\newblock \showarticletitle{Deeply fusing reviews and contents for cold start
  users in cross-domain recommendation systems}. In
  \bibinfo{booktitle}{\emph{Proceedings of the AAAI Conference on Artificial
  Intelligence}}, Vol.~\bibinfo{volume}{33}. \bibinfo{pages}{94--101}.
\newblock


\bibitem[\protect\citeauthoryear{Gabri{\'e}, Manoel, Luneau, Barbier, Macris,
  Krzakala, and Zdeborov{\'a}}{Gabri{\'e} et~al\mbox{.}}{2019}]%
        {gabrie2019entropy}
\bibfield{author}{\bibinfo{person}{Marylou Gabri{\'e}}, \bibinfo{person}{Andre
  Manoel}, \bibinfo{person}{Cl{\'e}ment Luneau}, \bibinfo{person}{Jean
  Barbier}, \bibinfo{person}{Nicolas Macris}, \bibinfo{person}{Florent
  Krzakala}, {and} \bibinfo{person}{Lenka Zdeborov{\'a}}.}
  \bibinfo{year}{2019}\natexlab{}.
\newblock \showarticletitle{Entropy and mutual information in models of deep
  neural networks}.
\newblock \bibinfo{journal}{\emph{Journal of Statistical Mechanics: Theory and
  Experiment}} \bibinfo{volume}{2019}, \bibinfo{number}{12}
  (\bibinfo{year}{2019}), \bibinfo{pages}{124014}.
\newblock


\bibitem[\protect\citeauthoryear{Ganin, Ustinova, Ajakan, Germain, Larochelle,
  Laviolette, Marchand, and Lempitsky}{Ganin et~al\mbox{.}}{2016}]%
        {ganin2016domain}
\bibfield{author}{\bibinfo{person}{Yaroslav Ganin}, \bibinfo{person}{Evgeniya
  Ustinova}, \bibinfo{person}{Hana Ajakan}, \bibinfo{person}{Pascal Germain},
  \bibinfo{person}{Hugo Larochelle}, \bibinfo{person}{Fran{\c{c}}ois
  Laviolette}, \bibinfo{person}{Mario Marchand}, {and} \bibinfo{person}{Victor
  Lempitsky}.} \bibinfo{year}{2016}\natexlab{}.
\newblock \showarticletitle{Domain-adversarial training of neural networks}.
\newblock \bibinfo{journal}{\emph{The Journal of Machine Learning Research}}
  \bibinfo{volume}{17}, \bibinfo{number}{1} (\bibinfo{year}{2016}),
  \bibinfo{pages}{2096--2030}.
\newblock


\bibitem[\protect\citeauthoryear{Gower, Loizou, Qian, Sailanbayev, Shulgin, and
  Richt{\'a}rik}{Gower et~al\mbox{.}}{2019}]%
        {gower2019sgd}
\bibfield{author}{\bibinfo{person}{Robert~Mansel Gower},
  \bibinfo{person}{Nicolas Loizou}, \bibinfo{person}{Xun Qian},
  \bibinfo{person}{Alibek Sailanbayev}, \bibinfo{person}{Egor Shulgin}, {and}
  \bibinfo{person}{Peter Richt{\'a}rik}.} \bibinfo{year}{2019}\natexlab{}.
\newblock \showarticletitle{SGD: General analysis and improved rates}. In
  \bibinfo{booktitle}{\emph{International Conference on Machine Learning}}.
  PMLR, \bibinfo{pages}{5200--5209}.
\newblock


\bibitem[\protect\citeauthoryear{Gretton, Bousquet, Smola, and
  Sch{\"o}lkopf}{Gretton et~al\mbox{.}}{2005}]%
        {gretton2005measuring}
\bibfield{author}{\bibinfo{person}{Arthur Gretton}, \bibinfo{person}{Olivier
  Bousquet}, \bibinfo{person}{Alex Smola}, {and} \bibinfo{person}{Bernhard
  Sch{\"o}lkopf}.} \bibinfo{year}{2005}\natexlab{}.
\newblock \showarticletitle{Measuring statistical dependence with
  Hilbert-Schmidt norms}. In \bibinfo{booktitle}{\emph{International conference
  on algorithmic learning theory}}. Springer, \bibinfo{pages}{63--77}.
\newblock


\bibitem[\protect\citeauthoryear{Guo, Tang, Chen, Zhu, Nguyen, and Yin}{Guo
  et~al\mbox{.}}{2021}]%
        {guo2021gcn}
\bibfield{author}{\bibinfo{person}{Lei Guo}, \bibinfo{person}{Li Tang},
  \bibinfo{person}{Tong Chen}, \bibinfo{person}{Lei Zhu}, \bibinfo{person}{Quoc
  Viet~Hung Nguyen}, {and} \bibinfo{person}{Hongzhi Yin}.}
  \bibinfo{year}{2021}\natexlab{}.
\newblock \showarticletitle{DA-GCN: A Domain-aware Attentive Graph Convolution
  Network for Shared-account Cross-domain Sequential Recommendation}.
\newblock \bibinfo{journal}{\emph{arXiv preprint arXiv:2105.03300}}
  (\bibinfo{year}{2021}).
\newblock


\bibitem[\protect\citeauthoryear{Hande, Puranik, Priyadharshini, and
  Chakravarthi}{Hande et~al\mbox{.}}{2021}]%
        {hande2021domain}
\bibfield{author}{\bibinfo{person}{Adeep Hande}, \bibinfo{person}{Karthik
  Puranik}, \bibinfo{person}{Ruba Priyadharshini}, {and}
  \bibinfo{person}{Bharathi~Raja Chakravarthi}.}
  \bibinfo{year}{2021}\natexlab{}.
\newblock \showarticletitle{Domain identification of scientific articles using
  transfer learning and ensembles}. In \bibinfo{booktitle}{\emph{Pacific-Asia
  Conference on Knowledge Discovery and Data Mining}}. Springer,
  \bibinfo{pages}{88--97}.
\newblock


\bibitem[\protect\citeauthoryear{He, Liao, Zhang, Nie, Hu, and Chua}{He
  et~al\mbox{.}}{2017}]%
        {he2017neural}
\bibfield{author}{\bibinfo{person}{Xiangnan He}, \bibinfo{person}{Lizi Liao},
  \bibinfo{person}{Hanwang Zhang}, \bibinfo{person}{Liqiang Nie},
  \bibinfo{person}{Xia Hu}, {and} \bibinfo{person}{Tat-Seng Chua}.}
  \bibinfo{year}{2017}\natexlab{}.
\newblock \showarticletitle{Neural collaborative filtering}. In
  \bibinfo{booktitle}{\emph{Proceedings of the 26th international conference on
  world wide web}}. \bibinfo{pages}{173--182}.
\newblock


\bibitem[\protect\citeauthoryear{Hu, Zhang, and Yang}{Hu et~al\mbox{.}}{2018}]%
        {hu2018conet}
\bibfield{author}{\bibinfo{person}{Guangneng Hu}, \bibinfo{person}{Yu Zhang},
  {and} \bibinfo{person}{Qiang Yang}.} \bibinfo{year}{2018}\natexlab{}.
\newblock \showarticletitle{Conet: Collaborative cross networks for
  cross-domain recommendation}. In \bibinfo{booktitle}{\emph{Proceedings of the
  27th ACM international conference on information and knowledge management}}.
  \bibinfo{pages}{667--676}.
\newblock


\bibitem[\protect\citeauthoryear{Kang, Jiang, Yang, and Hauptmann}{Kang
  et~al\mbox{.}}{2019b}]%
        {kang2019contrastive}
\bibfield{author}{\bibinfo{person}{Guoliang Kang}, \bibinfo{person}{Lu Jiang},
  \bibinfo{person}{Yi Yang}, {and} \bibinfo{person}{Alexander~G Hauptmann}.}
  \bibinfo{year}{2019}\natexlab{b}.
\newblock \showarticletitle{Contrastive adaptation network for unsupervised
  domain adaptation}. In \bibinfo{booktitle}{\emph{Proceedings of the IEEE/CVF
  Conference on Computer Vision and Pattern Recognition}}.
  \bibinfo{pages}{4893--4902}.
\newblock


\bibitem[\protect\citeauthoryear{Kang, Hwang, Lee, and Yu}{Kang
  et~al\mbox{.}}{2019a}]%
        {kang2019semi}
\bibfield{author}{\bibinfo{person}{SeongKu Kang}, \bibinfo{person}{Junyoung
  Hwang}, \bibinfo{person}{Dongha Lee}, {and} \bibinfo{person}{Hwanjo Yu}.}
  \bibinfo{year}{2019}\natexlab{a}.
\newblock \showarticletitle{Semi-supervised learning for cross-domain
  recommendation to cold-start users}. In \bibinfo{booktitle}{\emph{Proceedings
  of the 28th ACM International Conference on Information and Knowledge
  Management}}. \bibinfo{pages}{1563--1572}.
\newblock


\bibitem[\protect\citeauthoryear{Krishnan, Das, Bendre, Yang, and
  Sundaram}{Krishnan et~al\mbox{.}}{2020}]%
        {krishnan2020transfer}
\bibfield{author}{\bibinfo{person}{Adit Krishnan}, \bibinfo{person}{Mahashweta
  Das}, \bibinfo{person}{Mangesh Bendre}, \bibinfo{person}{Hao Yang}, {and}
  \bibinfo{person}{Hari Sundaram}.} \bibinfo{year}{2020}\natexlab{}.
\newblock \showarticletitle{Transfer Learning via Contextual Invariants for
  One-to-Many Cross-Domain Recommendation}. In
  \bibinfo{booktitle}{\emph{Proceedings of the 43rd International ACM SIGIR
  Conference on Research and Development in Information Retrieval}}.
  \bibinfo{pages}{1081--1090}.
\newblock


\bibitem[\protect\citeauthoryear{Li, Wang, Zhang, Li, Keutzer, Darrell, and
  Zhao}{Li et~al\mbox{.}}{2021}]%
        {li2021learning}
\bibfield{author}{\bibinfo{person}{Bo Li}, \bibinfo{person}{Yezhen Wang},
  \bibinfo{person}{Shanghang Zhang}, \bibinfo{person}{Dongsheng Li},
  \bibinfo{person}{Kurt Keutzer}, \bibinfo{person}{Trevor Darrell}, {and}
  \bibinfo{person}{Han Zhao}.} \bibinfo{year}{2021}\natexlab{}.
\newblock \showarticletitle{Learning invariant representations and risks for
  semi-supervised domain adaptation}. In \bibinfo{booktitle}{\emph{Proceedings
  of the IEEE/CVF Conference on Computer Vision and Pattern Recognition}}.
  \bibinfo{pages}{1104--1113}.
\newblock


\bibitem[\protect\citeauthoryear{Li, Chen, Ding, Zhu, Lu, and Shen}{Li
  et~al\mbox{.}}{2020}]%
        {li2020maximum}
\bibfield{author}{\bibinfo{person}{Jingjing Li}, \bibinfo{person}{Erpeng Chen},
  \bibinfo{person}{Zhengming Ding}, \bibinfo{person}{Lei Zhu},
  \bibinfo{person}{Ke Lu}, {and} \bibinfo{person}{Heng~Tao Shen}.}
  \bibinfo{year}{2020}\natexlab{}.
\newblock \showarticletitle{Maximum density divergence for domain adaptation}.
\newblock \bibinfo{journal}{\emph{IEEE transactions on pattern analysis and
  machine intelligence}} (\bibinfo{year}{2020}).
\newblock


\bibitem[\protect\citeauthoryear{Li and Tuzhilin}{Li and Tuzhilin}{2020}]%
        {li2020ddtcdr}
\bibfield{author}{\bibinfo{person}{Pan Li} {and} \bibinfo{person}{Alexander
  Tuzhilin}.} \bibinfo{year}{2020}\natexlab{}.
\newblock \showarticletitle{Ddtcdr: Deep dual transfer cross domain
  recommendation}. In \bibinfo{booktitle}{\emph{Proceedings of the 13th
  International Conference on Web Search and Data Mining}}.
  \bibinfo{pages}{331--339}.
\newblock


\bibitem[\protect\citeauthoryear{Li, Wang, Ren, Bing, and Lam}{Li
  et~al\mbox{.}}{2017}]%
        {li2017neural}
\bibfield{author}{\bibinfo{person}{Piji Li}, \bibinfo{person}{Zihao Wang},
  \bibinfo{person}{Zhaochun Ren}, \bibinfo{person}{Lidong Bing}, {and}
  \bibinfo{person}{Wai Lam}.} \bibinfo{year}{2017}\natexlab{}.
\newblock \showarticletitle{Neural rating regression with abstractive tips
  generation for recommendation}. In \bibinfo{booktitle}{\emph{Proceedings of
  the 40th International ACM SIGIR conference on Research and Development in
  Information Retrieval}}. \bibinfo{pages}{345--354}.
\newblock


\bibitem[\protect\citeauthoryear{Li, Tang, Li, and He}{Li
  et~al\mbox{.}}{2019}]%
        {li2019learning}
\bibfield{author}{\bibinfo{person}{Zejian Li}, \bibinfo{person}{Yongchuan
  Tang}, \bibinfo{person}{Wei Li}, {and} \bibinfo{person}{Yongxing He}.}
  \bibinfo{year}{2019}\natexlab{}.
\newblock \showarticletitle{Learning disentangled representation with pairwise
  independence}. In \bibinfo{booktitle}{\emph{Proceedings of the AAAI
  Conference on Artificial Intelligence}}, Vol.~\bibinfo{volume}{33}.
  \bibinfo{pages}{4245--4252}.
\newblock


\bibitem[\protect\citeauthoryear{Liu, Guo, Li, Zhao, and Wu}{Liu
  et~al\mbox{.}}{2021}]%
        {liu2021collaborative}
\bibfield{author}{\bibinfo{person}{Huiting Liu}, \bibinfo{person}{Lingling
  Guo}, \bibinfo{person}{Peipei Li}, \bibinfo{person}{Peng Zhao}, {and}
  \bibinfo{person}{Xindong Wu}.} \bibinfo{year}{2021}\natexlab{}.
\newblock \showarticletitle{Collaborative filtering with a deep adversarial and
  attention network for cross-domain recommendation}.
\newblock \bibinfo{journal}{\emph{Information Sciences}}  \bibinfo{volume}{565}
  (\bibinfo{year}{2021}), \bibinfo{pages}{370--389}.
\newblock


\bibitem[\protect\citeauthoryear{Man, Shen, Jin, and Cheng}{Man
  et~al\mbox{.}}{2017}]%
        {man2017cross}
\bibfield{author}{\bibinfo{person}{Tong Man}, \bibinfo{person}{Huawei Shen},
  \bibinfo{person}{Xiaolong Jin}, {and} \bibinfo{person}{Xueqi Cheng}.}
  \bibinfo{year}{2017}\natexlab{}.
\newblock \showarticletitle{Cross-Domain Recommendation: An Embedding and
  Mapping Approach.}. In \bibinfo{booktitle}{\emph{IJCAI}},
  Vol.~\bibinfo{volume}{17}. \bibinfo{pages}{2464--2470}.
\newblock


\bibitem[\protect\citeauthoryear{McAllester and Stratos}{McAllester and
  Stratos}{2020}]%
        {mcallester2020formal}
\bibfield{author}{\bibinfo{person}{David McAllester} {and}
  \bibinfo{person}{Karl Stratos}.} \bibinfo{year}{2020}\natexlab{}.
\newblock \showarticletitle{Formal limitations on the measurement of mutual
  information}. In \bibinfo{booktitle}{\emph{International Conference on
  Artificial Intelligence and Statistics}}. PMLR, \bibinfo{pages}{875--884}.
\newblock


\bibitem[\protect\citeauthoryear{McAuley and Leskovec}{McAuley and
  Leskovec}{2013}]%
        {mcauley2013hidden}
\bibfield{author}{\bibinfo{person}{Julian McAuley} {and} \bibinfo{person}{Jure
  Leskovec}.} \bibinfo{year}{2013}\natexlab{}.
\newblock \showarticletitle{Hidden factors and hidden topics: understanding
  rating dimensions with review text}. In \bibinfo{booktitle}{\emph{Proceedings
  of the 7th ACM conference on Recommender systems}}.
  \bibinfo{pages}{165--172}.
\newblock


\bibitem[\protect\citeauthoryear{Mikolov, Sutskever, Chen, Corrado, and
  Dean}{Mikolov et~al\mbox{.}}{2013}]%
        {mikolov2013distributed}
\bibfield{author}{\bibinfo{person}{Tomas Mikolov}, \bibinfo{person}{Ilya
  Sutskever}, \bibinfo{person}{Kai Chen}, \bibinfo{person}{Greg~S Corrado},
  {and} \bibinfo{person}{Jeff Dean}.} \bibinfo{year}{2013}\natexlab{}.
\newblock \showarticletitle{Distributed representations of words and phrases
  and their compositionality}.
\newblock \bibinfo{journal}{\emph{Advances in neural information processing
  systems}}  \bibinfo{volume}{26} (\bibinfo{year}{2013}).
\newblock


\bibitem[\protect\citeauthoryear{Mnih and Salakhutdinov}{Mnih and
  Salakhutdinov}{2008}]%
        {mnih2008probabilistic}
\bibfield{author}{\bibinfo{person}{Andriy Mnih} {and} \bibinfo{person}{Russ~R
  Salakhutdinov}.} \bibinfo{year}{2008}\natexlab{}.
\newblock \showarticletitle{Probabilistic matrix factorization}. In
  \bibinfo{booktitle}{\emph{Advances in neural information processing
  systems}}. \bibinfo{pages}{1257--1264}.
\newblock


\bibitem[\protect\citeauthoryear{Nema, Karatzoglou, and Radlinski}{Nema
  et~al\mbox{.}}{2021}]%
        {nema2021disentangling}
\bibfield{author}{\bibinfo{person}{Preksha Nema}, \bibinfo{person}{Alexandros
  Karatzoglou}, {and} \bibinfo{person}{Filip Radlinski}.}
  \bibinfo{year}{2021}\natexlab{}.
\newblock \showarticletitle{Disentangling Preference Representations for
  Recommendation Critiquing with {\ss}-VAE}. In
  \bibinfo{booktitle}{\emph{Proceedings of the 30th ACM International
  Conference on Information \& Knowledge Management}}.
  \bibinfo{pages}{1356--1365}.
\newblock


\bibitem[\protect\citeauthoryear{Panzeri, Senatore, Montemurro, and
  Petersen}{Panzeri et~al\mbox{.}}{2007}]%
        {panzeri2007correcting}
\bibfield{author}{\bibinfo{person}{Stefano Panzeri}, \bibinfo{person}{Riccardo
  Senatore}, \bibinfo{person}{Marcelo~A Montemurro}, {and}
  \bibinfo{person}{Rasmus~S Petersen}.} \bibinfo{year}{2007}\natexlab{}.
\newblock \showarticletitle{Correcting for the sampling bias problem in spike
  train information measures}.
\newblock \bibinfo{journal}{\emph{Journal of neurophysiology}}
  \bibinfo{volume}{98}, \bibinfo{number}{3} (\bibinfo{year}{2007}),
  \bibinfo{pages}{1064--1072}.
\newblock


\bibitem[\protect\citeauthoryear{Park, Lee, Yoo, Hur, and Yoon}{Park
  et~al\mbox{.}}{2020}]%
        {park2020joint}
\bibfield{author}{\bibinfo{person}{Changhwa Park}, \bibinfo{person}{Jonghyun
  Lee}, \bibinfo{person}{Jaeyoon Yoo}, \bibinfo{person}{Minhoe Hur}, {and}
  \bibinfo{person}{Sungroh Yoon}.} \bibinfo{year}{2020}\natexlab{}.
\newblock \showarticletitle{Joint contrastive learning for unsupervised domain
  adaptation}.
\newblock \bibinfo{journal}{\emph{arXiv preprint arXiv:2006.10297}}
  (\bibinfo{year}{2020}).
\newblock


\bibitem[\protect\citeauthoryear{Peng, Huang, Sun, and Saenko}{Peng
  et~al\mbox{.}}{2019}]%
        {peng2019domain}
\bibfield{author}{\bibinfo{person}{Xingchao Peng}, \bibinfo{person}{Zijun
  Huang}, \bibinfo{person}{Ximeng Sun}, {and} \bibinfo{person}{Kate Saenko}.}
  \bibinfo{year}{2019}\natexlab{}.
\newblock \showarticletitle{Domain agnostic learning with disentangled
  representations}. In \bibinfo{booktitle}{\emph{International Conference on
  Machine Learning}}. PMLR, \bibinfo{pages}{5102--5112}.
\newblock


\bibitem[\protect\citeauthoryear{Pennington, Socher, and Manning}{Pennington
  et~al\mbox{.}}{2014}]%
        {pennington2014glove}
\bibfield{author}{\bibinfo{person}{Jeffrey Pennington},
  \bibinfo{person}{Richard Socher}, {and} \bibinfo{person}{Christopher~D
  Manning}.} \bibinfo{year}{2014}\natexlab{}.
\newblock \showarticletitle{Glove: Global vectors for word representation}. In
  \bibinfo{booktitle}{\emph{Proceedings of the 2014 conference on empirical
  methods in natural language processing (EMNLP)}}.
  \bibinfo{pages}{1532--1543}.
\newblock


\bibitem[\protect\citeauthoryear{Ramakrishnan, Agrawal, and Lee}{Ramakrishnan
  et~al\mbox{.}}{2018}]%
        {ramakrishnan2018overcoming}
\bibfield{author}{\bibinfo{person}{Sainandan Ramakrishnan},
  \bibinfo{person}{Aishwarya Agrawal}, {and} \bibinfo{person}{Stefan Lee}.}
  \bibinfo{year}{2018}\natexlab{}.
\newblock \showarticletitle{Overcoming language priors in visual question
  answering with adversarial regularization}.
\newblock \bibinfo{journal}{\emph{arXiv preprint arXiv:1810.03649}}
  (\bibinfo{year}{2018}).
\newblock


\bibitem[\protect\citeauthoryear{Sachdeva and McAuley}{Sachdeva and
  McAuley}{2020}]%
        {sachdeva2020useful}
\bibfield{author}{\bibinfo{person}{Noveen Sachdeva} {and}
  \bibinfo{person}{Julian McAuley}.} \bibinfo{year}{2020}\natexlab{}.
\newblock \showarticletitle{How Useful are Reviews for Recommendation? A
  Critical Review and Potential Improvements}. In
  \bibinfo{booktitle}{\emph{Proceedings of the 43rd International ACM SIGIR
  Conference on Research and Development in Information Retrieval}}.
  \bibinfo{pages}{1845--1848}.
\newblock


\bibitem[\protect\citeauthoryear{Saito, Kim, Sclaroff, Darrell, and
  Saenko}{Saito et~al\mbox{.}}{2019}]%
        {saito2019semi}
\bibfield{author}{\bibinfo{person}{Kuniaki Saito}, \bibinfo{person}{Donghyun
  Kim}, \bibinfo{person}{Stan Sclaroff}, \bibinfo{person}{Trevor Darrell},
  {and} \bibinfo{person}{Kate Saenko}.} \bibinfo{year}{2019}\natexlab{}.
\newblock \showarticletitle{Semi-supervised domain adaptation via minimax
  entropy}. In \bibinfo{booktitle}{\emph{Proceedings of the IEEE/CVF
  International Conference on Computer Vision}}. \bibinfo{pages}{8050--8058}.
\newblock


\bibitem[\protect\citeauthoryear{Seo, Huang, Yang, and Liu}{Seo
  et~al\mbox{.}}{2017}]%
        {seo2017interpretable}
\bibfield{author}{\bibinfo{person}{Sungyong Seo}, \bibinfo{person}{Jing Huang},
  \bibinfo{person}{Hao Yang}, {and} \bibinfo{person}{Yan Liu}.}
  \bibinfo{year}{2017}\natexlab{}.
\newblock \showarticletitle{Interpretable convolutional neural networks with
  dual local and global attention for review rating prediction}. In
  \bibinfo{booktitle}{\emph{Proceedings of the eleventh ACM conference on
  recommender systems}}. \bibinfo{pages}{297--305}.
\newblock


\bibitem[\protect\citeauthoryear{Shi, Ma, Wang, Zhang, Fang, Xu, Liu, and
  Ma}{Shi et~al\mbox{.}}{2021}]%
        {shi2021wg4rec}
\bibfield{author}{\bibinfo{person}{Shaoyun Shi}, \bibinfo{person}{Weizhi Ma},
  \bibinfo{person}{Zhen Wang}, \bibinfo{person}{Min Zhang},
  \bibinfo{person}{Kun Fang}, \bibinfo{person}{Jingfang Xu},
  \bibinfo{person}{Yiqun Liu}, {and} \bibinfo{person}{Shaoping Ma}.}
  \bibinfo{year}{2021}\natexlab{}.
\newblock \showarticletitle{WG4Rec: Modeling Textual Content with Word Graph
  for News Recommendation}. In \bibinfo{booktitle}{\emph{Proceedings of the
  30th ACM International Conference on Information \& Knowledge Management}}.
  \bibinfo{pages}{1651--1660}.
\newblock


\bibitem[\protect\citeauthoryear{Shuai, Zhang, Wu, Sun, Hong, Wang, and
  Li}{Shuai et~al\mbox{.}}{2022}]%
        {shuai2022review}
\bibfield{author}{\bibinfo{person}{Jie Shuai}, \bibinfo{person}{Kun Zhang},
  \bibinfo{person}{Le Wu}, \bibinfo{person}{Peijie Sun},
  \bibinfo{person}{Richang Hong}, \bibinfo{person}{Meng Wang}, {and}
  \bibinfo{person}{Yong Li}.} \bibinfo{year}{2022}\natexlab{}.
\newblock \showarticletitle{A Review-aware Graph Contrastive Learning Framework
  for Recommendation}.
\newblock \bibinfo{journal}{\emph{arXiv preprint arXiv:2204.12063}}
  (\bibinfo{year}{2022}).
\newblock


\bibitem[\protect\citeauthoryear{Tan, Zhang, Liu, and Ma}{Tan
  et~al\mbox{.}}{2016}]%
        {tan2016rating}
\bibfield{author}{\bibinfo{person}{Yunzhi Tan}, \bibinfo{person}{Min Zhang},
  \bibinfo{person}{Yiqun Liu}, {and} \bibinfo{person}{Shaoping Ma}.}
  \bibinfo{year}{2016}\natexlab{}.
\newblock \showarticletitle{Rating-boosted latent topics: Understanding users
  and items with ratings and reviews.}. In \bibinfo{booktitle}{\emph{IJCAI}},
  Vol.~\bibinfo{volume}{16}. \bibinfo{pages}{2640--2646}.
\newblock


\bibitem[\protect\citeauthoryear{Tay, Luu, and Hui}{Tay et~al\mbox{.}}{2018}]%
        {tay2018multi}
\bibfield{author}{\bibinfo{person}{Yi Tay}, \bibinfo{person}{Anh~Tuan Luu},
  {and} \bibinfo{person}{Siu~Cheung Hui}.} \bibinfo{year}{2018}\natexlab{}.
\newblock \showarticletitle{Multi-pointer co-attention networks for
  recommendation}. In \bibinfo{booktitle}{\emph{Proceedings of the 24th ACM
  SIGKDD International Conference on Knowledge Discovery \& Data Mining}}.
  \bibinfo{pages}{2309--2318}.
\newblock


\bibitem[\protect\citeauthoryear{Wang, Ounis, and Macdonald}{Wang
  et~al\mbox{.}}{2021}]%
        {wang2021leveraging}
\bibfield{author}{\bibinfo{person}{Xi Wang}, \bibinfo{person}{Iadh Ounis},
  {and} \bibinfo{person}{Craig Macdonald}.} \bibinfo{year}{2021}\natexlab{}.
\newblock \showarticletitle{Leveraging Review Properties for Effective
  Recommendation}. In \bibinfo{booktitle}{\emph{Proceedings of the Web
  Conference 2021}}. \bibinfo{pages}{2209--2219}.
\newblock


\bibitem[\protect\citeauthoryear{Wang, Peng, Wang, Philip, Fu, and Hong}{Wang
  et~al\mbox{.}}{2018}]%
        {wang2018cross}
\bibfield{author}{\bibinfo{person}{Xinghua Wang}, \bibinfo{person}{Zhaohui
  Peng}, \bibinfo{person}{Senzhang Wang}, \bibinfo{person}{S~Yu Philip},
  \bibinfo{person}{Wenjing Fu}, {and} \bibinfo{person}{Xiaoguang Hong}.}
  \bibinfo{year}{2018}\natexlab{}.
\newblock \showarticletitle{Cross-domain recommendation for cold-start users
  via neighborhood based feature mapping}. In
  \bibinfo{booktitle}{\emph{International conference on database systems for
  advanced applications}}. Springer, \bibinfo{pages}{158--165}.
\newblock


\bibitem[\protect\citeauthoryear{Wu, Yang, Chen, Lian, Hong, and Wang}{Wu
  et~al\mbox{.}}{2020}]%
        {wu2020learning}
\bibfield{author}{\bibinfo{person}{Le Wu}, \bibinfo{person}{Yonghui Yang},
  \bibinfo{person}{Lei Chen}, \bibinfo{person}{Defu Lian},
  \bibinfo{person}{Richang Hong}, {and} \bibinfo{person}{Meng Wang}.}
  \bibinfo{year}{2020}\natexlab{}.
\newblock \showarticletitle{Learning to transfer graph embeddings for inductive
  graph based recommendation}. In \bibinfo{booktitle}{\emph{Proceedings of the
  43rd International ACM SIGIR Conference on Research and Development in
  Information Retrieval}}. \bibinfo{pages}{1211--1220}.
\newblock


\bibitem[\protect\citeauthoryear{Xiong, Ye, Chen, Zhang, Zhao, Hu, Zhang, and
  Zhou}{Xiong et~al\mbox{.}}{2021}]%
        {xiong2021counterfactual}
\bibfield{author}{\bibinfo{person}{Kun Xiong}, \bibinfo{person}{Wenwen Ye},
  \bibinfo{person}{Xu Chen}, \bibinfo{person}{Yongfeng Zhang},
  \bibinfo{person}{Wayne~Xin Zhao}, \bibinfo{person}{Binbin Hu},
  \bibinfo{person}{Zhiqiang Zhang}, {and} \bibinfo{person}{Jun Zhou}.}
  \bibinfo{year}{2021}\natexlab{}.
\newblock \showarticletitle{Counterfactual Review-based Recommendation}. In
  \bibinfo{booktitle}{\emph{Proceedings of the 30th ACM International
  Conference on Information \& Knowledge Management}}.
  \bibinfo{pages}{2231--2240}.
\newblock


\bibitem[\protect\citeauthoryear{Yang, Liu, Su, Tang, Liu, and He}{Yang
  et~al\mbox{.}}{2021}]%
        {yang2021autoft}
\bibfield{author}{\bibinfo{person}{Xiangli Yang}, \bibinfo{person}{Qing Liu},
  \bibinfo{person}{Rong Su}, \bibinfo{person}{Ruiming Tang},
  \bibinfo{person}{Zhirong Liu}, {and} \bibinfo{person}{Xiuqiang He}.}
  \bibinfo{year}{2021}\natexlab{}.
\newblock \showarticletitle{AutoFT: Automatic Fine-Tune for Parameters Transfer
  Learning in Click-Through Rate Prediction}.
\newblock \bibinfo{journal}{\emph{arXiv preprint arXiv:2106.04873}}
  (\bibinfo{year}{2021}).
\newblock


\bibitem[\protect\citeauthoryear{Yu, Lin, Ge, Ou, and Qin}{Yu
  et~al\mbox{.}}{2020}]%
        {yu2020semi}
\bibfield{author}{\bibinfo{person}{Wenhui Yu}, \bibinfo{person}{Xiao Lin},
  \bibinfo{person}{Junfeng Ge}, \bibinfo{person}{Wenwu Ou}, {and}
  \bibinfo{person}{Zheng Qin}.} \bibinfo{year}{2020}\natexlab{}.
\newblock \showarticletitle{Semi-supervised collaborative filtering by
  text-enhanced domain adaptation}. In \bibinfo{booktitle}{\emph{Proceedings of
  the 26th ACM SIGKDD International Conference on Knowledge Discovery \& Data
  Mining}}. \bibinfo{pages}{2136--2144}.
\newblock


\bibitem[\protect\citeauthoryear{Yu, Zhang, He, Chen, Xiong, and Qin}{Yu
  et~al\mbox{.}}{2018}]%
        {yu2018aesthetic}
\bibfield{author}{\bibinfo{person}{Wenhui Yu}, \bibinfo{person}{Huidi Zhang},
  \bibinfo{person}{Xiangnan He}, \bibinfo{person}{Xu Chen}, \bibinfo{person}{Li
  Xiong}, {and} \bibinfo{person}{Zheng Qin}.} \bibinfo{year}{2018}\natexlab{}.
\newblock \showarticletitle{Aesthetic-based clothing recommendation}. In
  \bibinfo{booktitle}{\emph{Proceedings of the 2018 world wide web
  conference}}. \bibinfo{pages}{649--658}.
\newblock


\bibitem[\protect\citeauthoryear{Yuan, He, Karatzoglou, and Zhang}{Yuan
  et~al\mbox{.}}{2020}]%
        {yuan2020parameter}
\bibfield{author}{\bibinfo{person}{Fajie Yuan}, \bibinfo{person}{Xiangnan He},
  \bibinfo{person}{Alexandros Karatzoglou}, {and} \bibinfo{person}{Liguang
  Zhang}.} \bibinfo{year}{2020}\natexlab{}.
\newblock \showarticletitle{Parameter-efficient transfer from sequential
  behaviors for user modeling and recommendation}. In
  \bibinfo{booktitle}{\emph{Proceedings of the 43rd International ACM SIGIR
  Conference on Research and Development in Information Retrieval}}.
  \bibinfo{pages}{1469--1478}.
\newblock


\bibitem[\protect\citeauthoryear{Yuan, Yao, and Benatallah}{Yuan
  et~al\mbox{.}}{2019}]%
        {yuan2019darec}
\bibfield{author}{\bibinfo{person}{Feng Yuan}, \bibinfo{person}{Lina Yao},
  {and} \bibinfo{person}{Boualem Benatallah}.} \bibinfo{year}{2019}\natexlab{}.
\newblock \showarticletitle{DARec: deep domain adaptation for cross-domain
  recommendation via transferring rating patterns}.
\newblock \bibinfo{journal}{\emph{arXiv preprint arXiv:1905.10760}}
  (\bibinfo{year}{2019}).
\newblock


\bibitem[\protect\citeauthoryear{Zeng, Xu, and Ai}{Zeng et~al\mbox{.}}{2021}]%
        {zeng2021zero}
\bibfield{author}{\bibinfo{person}{Hansi Zeng}, \bibinfo{person}{Zhichao Xu},
  {and} \bibinfo{person}{Qingyao Ai}.} \bibinfo{year}{2021}\natexlab{}.
\newblock \showarticletitle{A Zero Attentive Relevance Matching Networkfor
  Review Modeling in Recommendation System}.
\newblock \bibinfo{journal}{\emph{arXiv preprint arXiv:2101.06387}}
  (\bibinfo{year}{2021}).
\newblock


\bibitem[\protect\citeauthoryear{Zhao, Li, Xiao, Deng, and Sun}{Zhao
  et~al\mbox{.}}{2020}]%
        {zhao2020catn}
\bibfield{author}{\bibinfo{person}{Cheng Zhao}, \bibinfo{person}{Chenliang Li},
  \bibinfo{person}{Rong Xiao}, \bibinfo{person}{Hongbo Deng}, {and}
  \bibinfo{person}{Aixin Sun}.} \bibinfo{year}{2020}\natexlab{}.
\newblock \showarticletitle{CATN: Cross-Domain Recommendation for Cold-Start
  Users via Aspect Transfer Network}.
\newblock \bibinfo{journal}{\emph{arXiv preprint arXiv:2005.10549}}
  (\bibinfo{year}{2020}).
\newblock


\bibitem[\protect\citeauthoryear{Zhao, Ma, Chen, and Deng}{Zhao
  et~al\mbox{.}}{2021}]%
        {zhao2021domain}
\bibfield{author}{\bibinfo{person}{Haiteng Zhao}, \bibinfo{person}{Chang Ma},
  \bibinfo{person}{Qinyu Chen}, {and} \bibinfo{person}{Zhihong Deng}.}
  \bibinfo{year}{2021}\natexlab{}.
\newblock \showarticletitle{Domain Adaptation via Maximizing Surrogate Mutual
  Information}.
\newblock \bibinfo{journal}{\emph{arXiv preprint arXiv:2110.12184}}
  (\bibinfo{year}{2021}).
\newblock


\bibitem[\protect\citeauthoryear{Zheng, Noroozi, and Yu}{Zheng
  et~al\mbox{.}}{2017}]%
        {zheng2017joint}
\bibfield{author}{\bibinfo{person}{Lei Zheng}, \bibinfo{person}{Vahid Noroozi},
  {and} \bibinfo{person}{Philip~S Yu}.} \bibinfo{year}{2017}\natexlab{}.
\newblock \showarticletitle{Joint deep modeling of users and items using
  reviews for recommendation}. In \bibinfo{booktitle}{\emph{Proceedings of the
  Tenth ACM International Conference on Web Search and Data Mining}}.
  \bibinfo{pages}{425--434}.
\newblock


\bibitem[\protect\citeauthoryear{Zhu, Wang, Chen, Zhou, Li, and Liu}{Zhu
  et~al\mbox{.}}{2021}]%
        {zhu2021cross}
\bibfield{author}{\bibinfo{person}{Feng Zhu}, \bibinfo{person}{Yan Wang},
  \bibinfo{person}{Chaochao Chen}, \bibinfo{person}{Jun Zhou},
  \bibinfo{person}{Longfei Li}, {and} \bibinfo{person}{Guanfeng Liu}.}
  \bibinfo{year}{2021}\natexlab{}.
\newblock \showarticletitle{Cross-domain recommendation: challenges, progress,
  and prospects}.
\newblock \bibinfo{journal}{\emph{arXiv preprint arXiv:2103.01696}}
  (\bibinfo{year}{2021}).
\newblock


\end{thebibliography}

\end{document}